\documentclass[aps,prb,superscriptaddress,twocolumn,showpacs,amsmath,amssymb,groupedaddress]{revtex4}

\usepackage{color}
\usepackage{graphicx}
\usepackage{bm}
\definecolor{Blue}{rgb}{0.00, 0.00, 1.00}
\definecolor{Red}{rgb}{1.00, 0.00, 0.00}

\begin{document}

\title{Quantum Hall effect and semiconductor to semimetal transition in biased black phosphorus}

\author{Shengjun Yuan}
\email{s.yuan@science.ru.nl }
\affiliation{Radboud University, Institute for Molecules and Materials, Heijendaalseweg 135, NL-6525 AJ Nijmegen, The Netherlands}
\author{Edo van Veen}
\affiliation{Radboud University, Institute for Molecules and Materials, Heijendaalseweg 135, NL-6525 AJ Nijmegen, The Netherlands}
\author{Mikhail I. Katsnelson}
\affiliation{Radboud University, Institute for Molecules and Materials, Heijendaalseweg 135, NL-6525 AJ Nijmegen, The Netherlands}
\author{ Rafael Rold\'an}
\email{rroldan@icmm.csic.es}
\affiliation{Instituto de Ciencia de Materiales de Madrid, CSIC, Cantoblanco E28049 Madrid, Spain}

\date{\today}

\begin{abstract}
We study the quantum Hall effect of 2D electron gas in black phosphorus in the presence of perpendicular electric and magnetic fields. In the absence of a bias voltage, the external magnetic field leads to a quantization of the energy spectrum into equidistant Landau levels, with different cyclotron frequencies for the electron and hole bands. The applied voltage reduces the  band gap, and eventually a semiconductor to semimetal transition takes place. This nontrivial phase is characterized by the emergence of a pair of Dirac points in the spectrum. As a consequence, the Landau levels are not equidistant anymore, but follow the $\varepsilon_n\propto \sqrt{nB}$ characteristic of Dirac crystals as graphene. By using the Kubo-Bastin formula in the context of the kernel polynomial method, we compute the Hall conductivity of the system. We obtain a $\sigma_{xy}\propto 2n$ quantization of the Hall conductivity in the gapped phase (standard quantum Hall effect regime), and a $\sigma_{xy}\propto 4(n+1/2)$ quantization in the semimetalic phase, characteristic of Dirac systems with non-trivial topology.

\end{abstract}


\maketitle

\section{Introduction}

Black phosphorus (BP) is a direct band gap semiconductor that has been recently exfoliated to obtain atomically thin samples.\cite{li14,xia1,C15} Each BP layer forms a puckered surface due to $sp^3$ hybridization, revealing a highly anisotropic electrical mobility,\cite{morita86,liu14} ambipolar field effect, linear dichroism in optical absorption spectra,\cite{morita86,qiao14,low14cond,xia1,YRK15} and anisotropic plasmons.\cite{low14plas}
Encapsulation of BP with hexagonal boron nitride (hBN) has lead to high carrier mobility devices, with the observation of quantum magneto-oscillations\cite{exp1,exp2,exp3,exp4,cao15} and integer quantum Hall effect.\cite{Li2015} One of the most surprising characteristics of BP is its strong response to external electric and strain fields. As a consequence, the electronic and optical properties of this material can be efficiently tuned by applying an external bias voltage\cite{KK15,LZ15,DQ15,Chaves2015,Baik_2015} or by strain engineering.\cite{Rodin2014,RG15,XC15,QC15,Manjanath2015}  In particular, it is possible to drive a semiconductor to semimetal transition, with the appearance of Dirac like dispersion.\cite{KK15,LZ15,DQ15,JRKY15}

In this paper we study the electronic spectrum of biased BP in the presence of a strong magnetic field. For this we use a tight-binding
model which properly accounts for the band structure in a wide
energy window of the spectrum.\cite{Rudenko2014,Rudenko2015} The electronic density of states (DOS) is calculated from the solution of the time-dependent Schr\"odinger equation within the framework of  the tight-binding propagation method (TBPM),\cite%
{YRK10,Yuan2011,Yuan2012} which is an efficient numerical tool in large-scale
calculations of realistic systems with more than millions of atoms. In the absence or for moderate values of the applied bias, the obtained Landau level quantization is that of a standard two dimensional electron gas (2DEG) with a set of equidistant Landau levels.\cite{Pereira15,Jiang2015,zhou15} When the applied electric field is strong enough, the BP suffers a semiconductor to semimetal transition, with the appearance of a set of non-equidistant LLs, associated to Dirac like cones that emerge in the spectrum. Such a LL spectrum resembles that of graphene in the quantum Hall regime, with the difference that biased BP presents a pronounced electron-hole asymmetry. As we increase the energy, the spectrum acquires a highly nontrivial quantization due to the presence of a Van Hove singularity, with a corresponding change in the topological Berry phase. We further calculated the Hall conductivity from the Kubo-Bastin formula,\cite{Bastin71} in the context of the kernel polynomial method.\cite{GCR15,Ortmann2015} We find that unbiased BP presents the characteristic integer quantum Hall effect with $\sigma^{{\rm IQHE}}_{xy}=2n(e^2/h)$, whereas biased semimetal BP presents a {\it relativistic} quantum Hall effect characteristic of Dirac materials, with $\sigma^{{\rm RQHE}}_{xy}=4(n+1/2)(e^2/h)$.\cite{G11} Although we perform the numerical calculations for the simplest case of bilayer BP, the physical results should hold for any multilayer sample exposed to external magnetic and electric fields. 

\begin{figure}[b]
\centering
\includegraphics[width=0.45\textwidth]{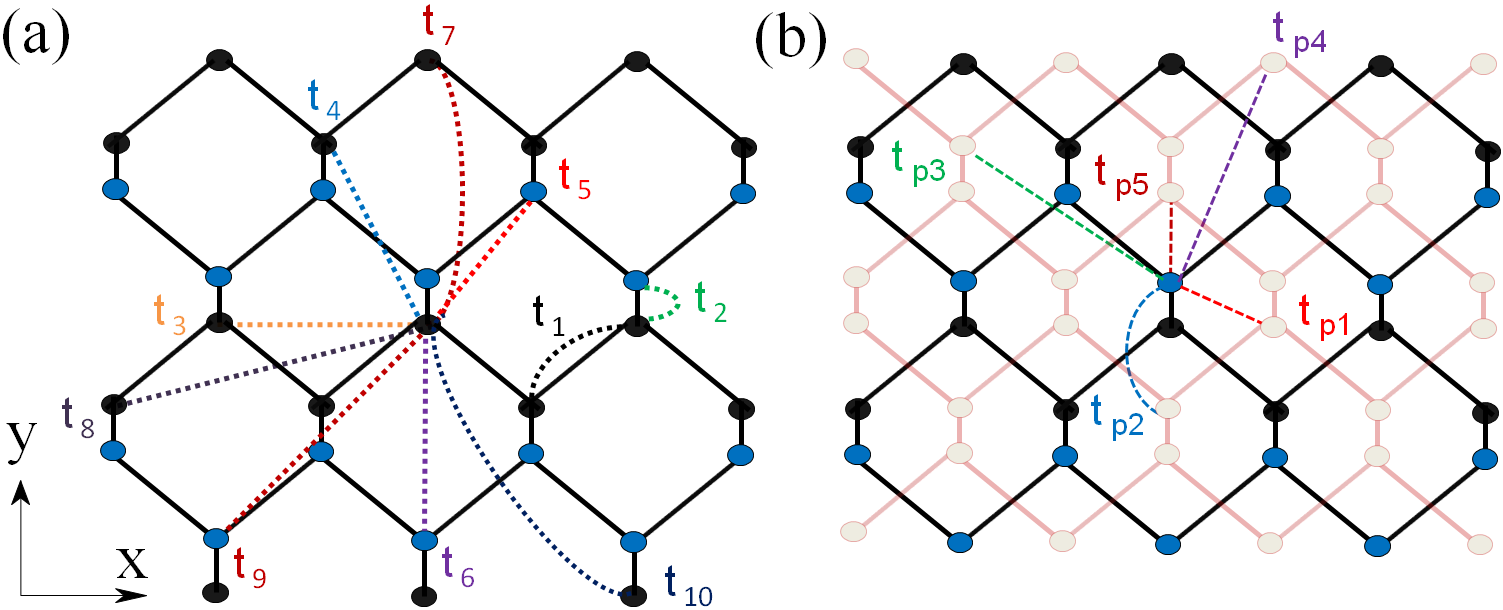}
\caption{Lattice structure of single layer (a) and bilayer (b) black phosphorus. Circles of different color correspond to atoms located in different planes within a single puckered layer.  The relevant hopping terms considered in the Hamiltonian (\ref{Eq:Hamiltonian})
 are indicated: 10 in-plane hopping terms (a), and 5 inter-layer terms (b). }
  \label{Fig:Lattice}
\end{figure}

\begin{figure*}[t]
\centering
\includegraphics[width=0.7\textwidth]{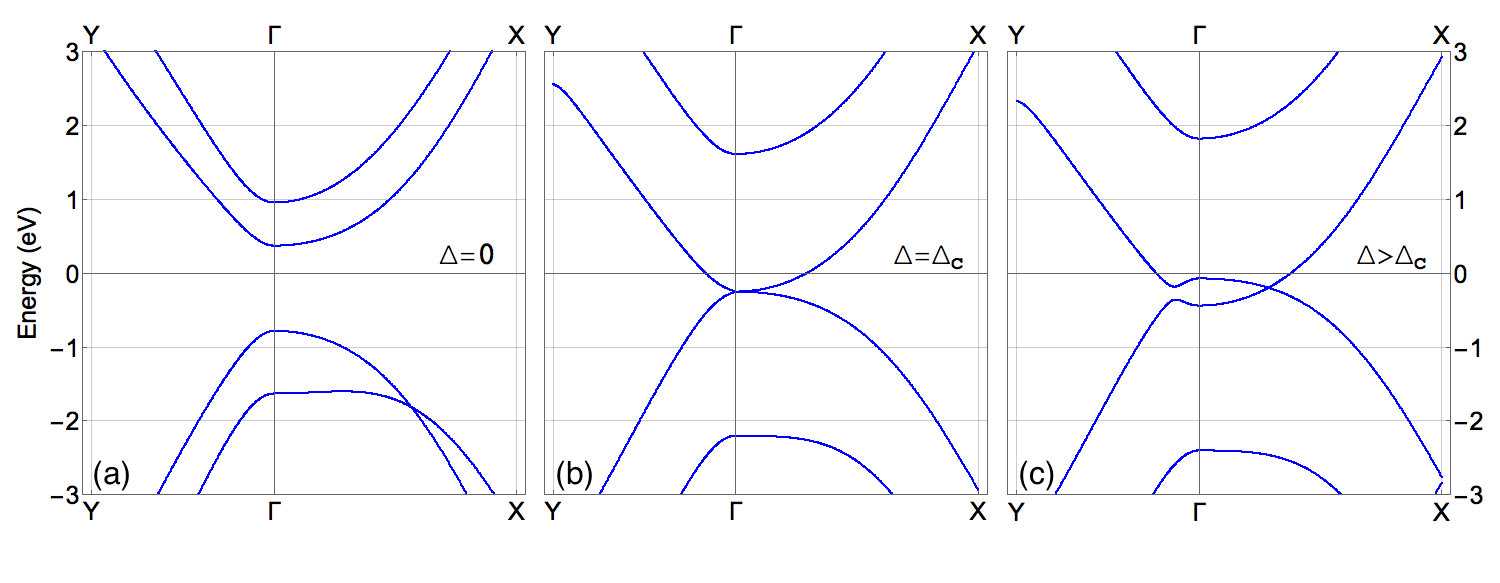}
\caption{Band structure of biased bilayer BP for three representative values of the applied voltage: unbiased ($\Delta=0$) for which the system is gapped, $\Delta=\Delta_c$ for which the gap closes and there is a band crossing at the $\Gamma$ point, and $\Delta>\Delta_c$ corresponding to the semimetal phase with the creation of Dirac points in the $\Gamma-{\rm X}$ direction of the Brillouin zone.}
  \label{Fig:Bands}
\end{figure*}

\section{Electronic band structure and Landau quantization}

\begin{figure}[b]
\centering
\includegraphics[width=0.35\textwidth]{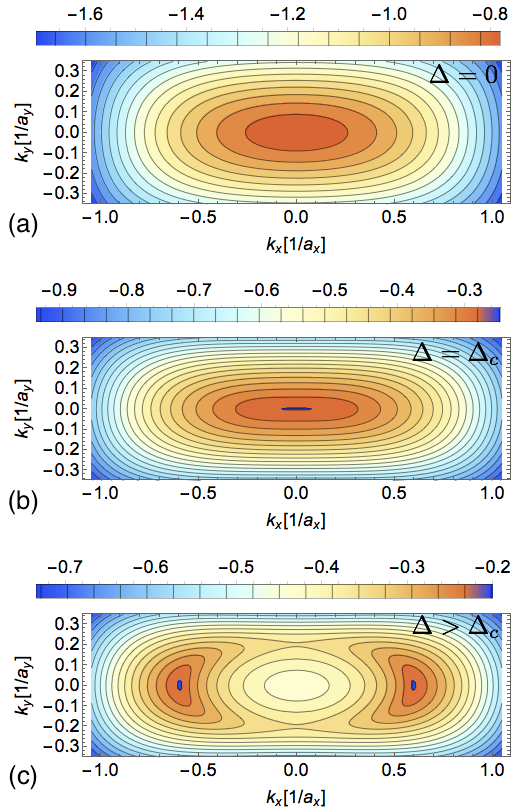}
\caption{Constant energy contours of biased bilayer BP for the three values of the applied voltage used in Fig. \ref{Fig:Bands}. For $\Delta>\Delta_c$ corresponding to the semimetal phase with the creation of Dirac points in the $\Gamma-{\rm X}$ direction of the Brillouin zone.}
  \label{Fig:CECs}
\end{figure}

BP is formed by stacking of phosphorene layers, coupled by Van der Waals interaction. 
Single layer BP contains two atomic layers and two kinds of P-P bonds (in-plane and inter-plane),\cite{morita86} as shown in Fig. \ref{Fig:Lattice}.
Our calculations are done using a $GW$ based tight-binding model that properly reproduces 
the conduction and valence bands in an energy range of $\sim $%
0.3 eV beyond the gap\cite{Rudenko2014,Rudenko2015} 
\begin{equation}
\mathcal{H}=\sum_{i\neq j}t_{ij}c_{i}^{\dagger
}c_{j}+\sum_{i\neq j}t_{p,ij}c_{i}^{\dagger }c_{j},  \label{Eq:Hamiltonian}
\end{equation}%
where $c_{i}^{\dagger}$ ($c_{i}$) creates (annihilates) an electron at site $i$, and ten intra-layer $t_{ij}$ and five inter-layer $t_{p,ij}$ hopping terms are considered in the model.
The values of the ten intra-layer hopping terms [shown in Fig. \ref{Fig:Lattice}(a)] are $t_{1}=-1.486$ eV, $t_{2}=3.729$ eV, $t_{3}=-0.252$ eV, $t_{4}=-0.071$ eV, $t_{5}=-0.019$ eV, $t_{6}=-0.186$ eV, $t_{7}=-0.063$ eV, $t_{8}=0.101$ eV, $t_{9}=-0.042$ eV, $t_{10}=-0.073$ eV, and the five inter-layer hopping terms [Fig.\ref{Fig:Lattice}(b)] are $t_{p1}=0.524$ eV, $t_{p2}=0.180$ eV, $t_{p3}=-0.123$ eV, $t_{p4}=-0.168$ eV, $t_{p5}=0.005$ eV. \cite{Rudenko2015} 
The effect of an electric field on the electronic dispersion is considered by introducing linearly a biased on-site potential 
difference between the out-most planes of two layers, without considering the screening effect. For example, in a single layer we include a different on-site potential $\pm \Delta/2$ in the top and bottom sublayers, respectively, whereas in a bilayer BP we include a sequence of on-site potentials in the four planes with the form $\Delta/2+v_b\Delta$, $\Delta/2-v_b\Delta$, $-\Delta/2+v_b\Delta$ and $-\Delta/2-v_b\Delta$, where $v_b=0.202$ is a linear scaling factor accounting for the lattice position along the direction of the external electric field.\cite{Rudenko2015} Fig. \ref{Fig:Bands} show the band structure obtained from the tight-binding model (\ref{Eq:Hamiltonian}) for three representative cases, and their corresponding constant energy contours (CEC) are shown in Fig. \ref{Fig:CECs}. As it is well known,\cite{morita86} for unbiased BP ($\Delta=0$) the band structure corresponds to an anisotropic direct band gap semiconductor, with the gap placed at the $\Gamma$ point of the Brillouin zone. 

\begin{figure}[t]
\centering
\includegraphics[width=0.23\textwidth]{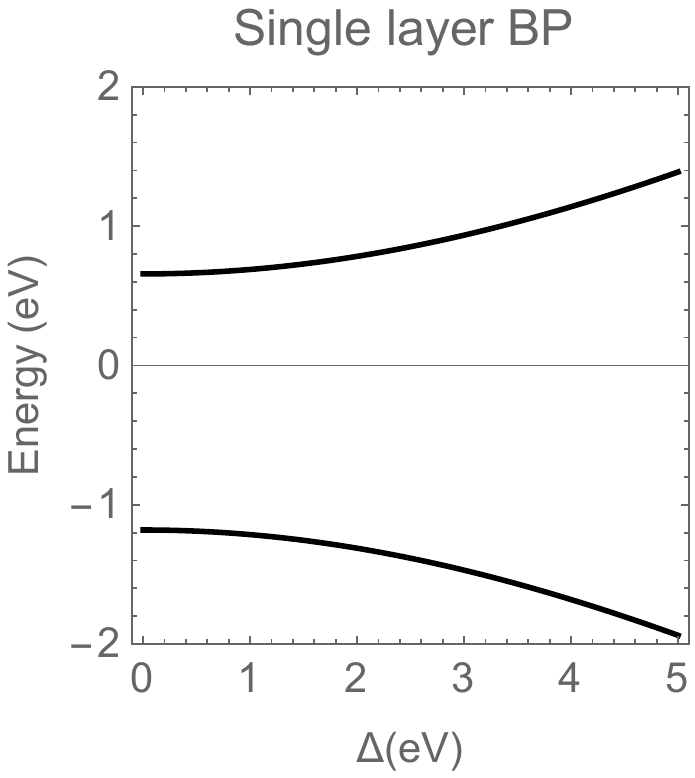}
\includegraphics[width=0.23\textwidth]{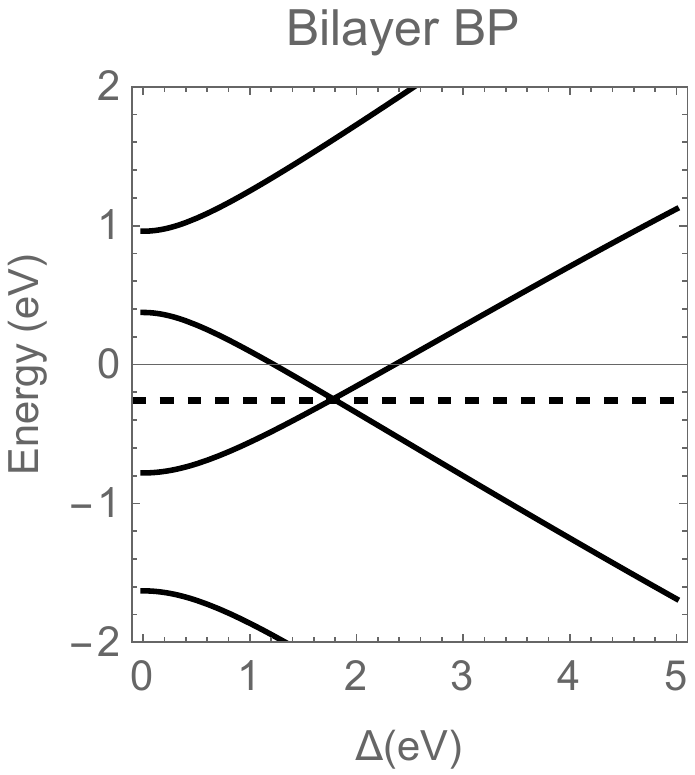}
\caption{Evolution of the valence and conduction band edges at the $\Gamma$ point as a function of the biased potential $\Delta$, for single layer and bilayer BP, as calculated from the full tight-binding model (\ref{Eq:Hamiltonian}). The dashed line in panel (b) indicates the chemical potential energy in the semimetal phase.}
  \label{Fig:GapU}
\end{figure}

It is interesting to consider the different effect of a perpendicular electric field in the band structure of single layer and multilayer BP. In Fig. \ref{Fig:GapU} we show evolution of the band gap at the $\Gamma$ point as a function of the biased potential, defined from the energy difference between the valence and conduction band edges as obtained from the full tight-binding model (\ref{Eq:Hamiltonian}). We observe that, whereas the gap increases with $\Delta$ in single layer BP, the gap in bilayer BP decreases with the applied bias, and eventually a semiconductor to semimetal transition takes place. A similar closing of the gap with the bias potential occurs for any multilayer sample. The opposite behavior between single layer and multilayer can be understood analytically by using the tight-binding model (\ref{Eq:Hamiltonian}) with only the leading hopping terms, namely $t_1$, $t_2$ and $t_{p1}$. In the absence of a perpendicular electric field, $\Delta=0$, the gap in single layer is controlled by the difference between the in-plane hopping parameter $t_1$ and the inter-plane $t_2$, 
\begin{equation}\label{Eq:Gap1LDelta0}
E_{g1L}(\Delta=0)\approx 4t_1+2t_2
\end{equation} where it is important to notice the different sign of the two terms, $t_1\approx-1.5~{\rm eV}<0$ and $t_2\approx3.7~{\rm eV}>0$. For bilayer BP, the gap at $\Delta=0$ is approximately given by
\begin{equation}
E_{g2L}(\Delta=0)\approx 4t_1+2\sqrt{t_2^2+2t_{p1}^2-2|t_{p1}|\sqrt{t_2^2+t_{p1}^2}}.
\end{equation} 
Notice that the inter-layer hopping $t_{p1}$ in a bilayer BP enters in the gap equation as an extra contribution to the inter-plane hopping term in single-layer BP, $t_2$.
In the presence of a biased potential, and within the above approximation, the gap in single layer BP can be expressed as 
\begin{equation}\label{Eq:Gap1L}
E_{g1L}\approx 4t_1+2\sqrt{t_2^2+\left(\frac{\Delta}{2}\right)^2},
\end{equation}
whereas for bilayer BP the gap is given by
\begin{equation}\label{Eq:Gap2L}
E_{g2L}\approx 4t_1+2\sqrt{t_2^2+\left(\frac{\Delta}{2}\right)^2+f_{{\rm inter}}},
\end{equation}
where we have defined
\begin{equation}\label{Eq:finter}
f_{{\rm inter}}=2t_{p1}^2+v_b^2\Delta^2-\sqrt{t_2^2(4t_{p1}^2+\Delta^2)+(-2t_{p1}^2+v_b\Delta^2)^2}.
\end{equation}
One can easily see that, for the hopping parameters of the model, there is no real solution for $\Delta$ that {\it closes} the gap in single layer BP, which should fulfill
$
\Delta^c_{1L}\approx 2\sqrt{4t_1^2-t_2^2}.
$
Therefore, this simple analytical analysis shows that application of a perpendicular electric field has the effect of opening the gap in single layer BP, in agreement with the full tight-binding results shown in Fig. \ref{Fig:GapU}. Bilayer BP has, on the other hand, a real solution for the closing of the gap. The analytical expression is too long as to be given here, but one can simply observe that  the term $f_{\rm inter}$, as defined in Eq.(\ref{Eq:finter}), is $<0$. This leads to a correction for the second contribution in the gap equation (\ref{Eq:Gap2L}), which can fully cancel the $4t_1$ term, driving a semiconductor to semimetal transition. From now on, we will focus on the multilayer case, for which the aforementioned transition can take place in the presence of a bias potential. The topological nature of the transition has been addressed, by combining DFT and group theory analysis, by Liu et al.\cite{LZ15}. Semiconducting unbiased black phosphorus have valence and conduction bands with different symmetric representations at the $\Gamma$ point (point group $D_{2h}$): conduction band has the representation $A_g (\Gamma_{1})$, whereas valence band has representation $B_{3u} (\Gamma_{8})$. One can define the inversion energy as $\Delta_{inv}=E_{\Gamma 1} - E_{\Gamma 8}$. When the bias voltage is large enough, the gap is zero and $\Delta_{inv}$ becomes negative, indicating a band inversion. This band inversion is accompanied by a Dirac-like band crossing, as seen in Fig. \ref{Fig:Bands}. This band crossing is protected by fractional translation symmetry due to the different character of the two bands. Therefore, the spectrum can be described at low energies by a $2\times2$ Dirac equation. The analysis of the wave-function performed in Ref. \onlinecite{LZ15} for multilayer samples reveals that the $\Gamma_{1}$ states are mainly localised in the top layer, whereas the spectral weight of the $\Gamma_8$ states is stronger in the bottom layer.

We insist that the approximation considered here does not take into account electrostatic screening due to the external electric field. This effect has been studied, using a low energy continuum model and within a nonlinear Thomas-Fermi theory, in Ref. \onlinecite{low14plas}. The potential difference across a BP sample obtained there suggests that black phosphorus presents an intermediate  screening behaviour between the strong coupling limit of graphene, where the carriers concentrate close to the interface, and the weak coupling regime with reduced screening properties that dominates the screening of other  van der Waals semiconducting materials as MoS$_2$.

\begin{figure}[t]
\centering
 \includegraphics[width=0.45\textwidth]{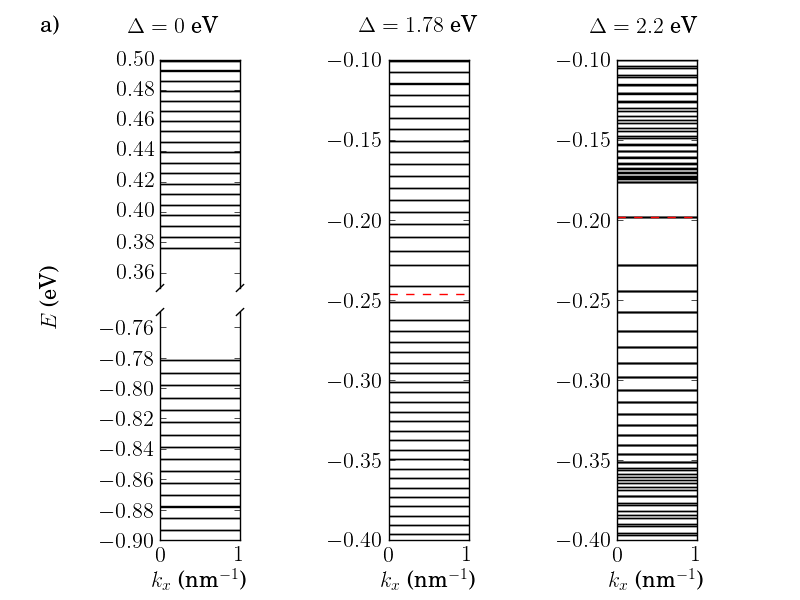}
 \includegraphics[width=0.45\textwidth]{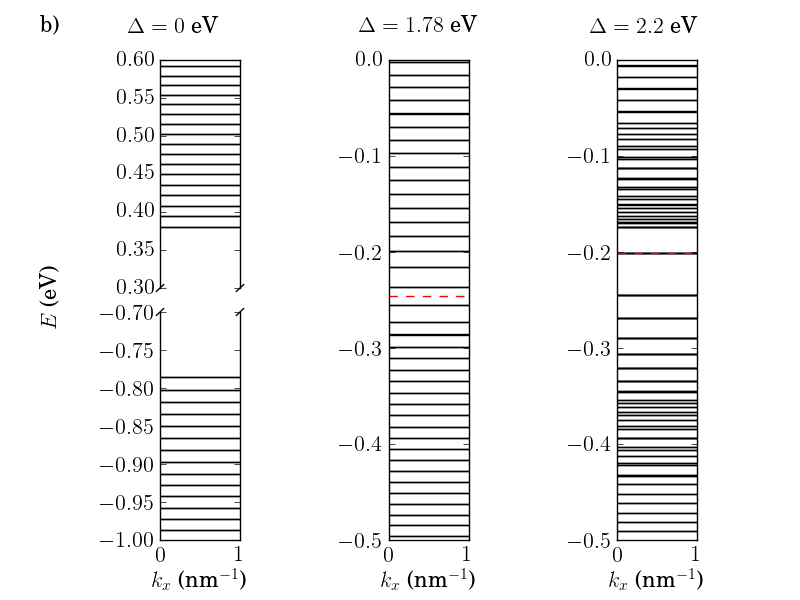}
  \includegraphics[width=0.45\textwidth]{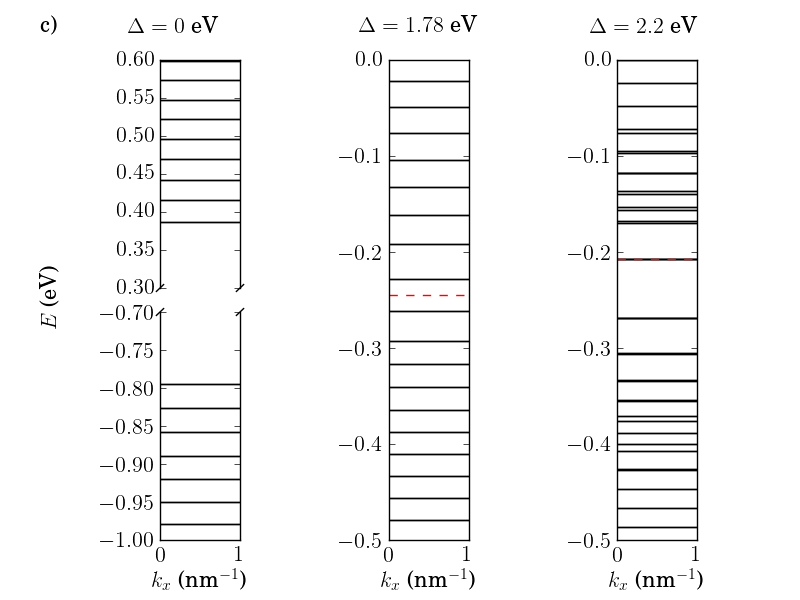}
\caption{Landau level structure obtained from exact diagonalization of the Hamiltonian (\ref{Eq:Hamiltonian-k}) for a) $B=32.5$~T, b) $B=65$~T and  a) $B=130$~T. The Fermi level is indicated by a red dashed line, and falls in the gap (skipped  regions) for $\Delta = 0$.}
  \label{Fig:LLs-Diagonalization}
\end{figure}

The presence of a magnetic field is accounted by means of the Peierls substitution, which replace the hopping term between two sites to 
\begin{equation}
t_{ij}\rightarrow t_{ij}\exp{\left[i\frac{2\pi}{\Phi_0}e\int_{{\bf R}_i}^{{\bf R}_j} {\bf A}\cdot d{\bf l}\right]},
\end{equation} 
where $\Phi_0=hc/e$ is the flux quantum and the vector potential in the Landau gauge is ${\bf A}=(-By,0,0)$, $B$ being the strength of the magnetic field.  The band structure can now be calculated by choosing a ribbon with one unit cell width and a height that exactly matches the period of the Peierls phase. After obtaining the Hamiltonian as a function of momentum
\begin{equation}\label{Eq:Hamiltonian-k}
\mathcal{H}(\mathbf{k}) =\sum_{i\neq j}t_{ij}c_{i}^{\dagger
}c_{j}e^{i \mathbf{k} \cdot (\mathbf{r}_i - \mathbf{r}_j)} +\sum_{i\neq j}t_{p,ij}c_{i}^{\dagger }c_{j} e^{i \mathbf{k} \cdot (\mathbf{r}_i - \mathbf{r}_j)},
\end{equation} 
the energy eigenvalues corresponding to a momentum $\mathbf{k}$ can be found with exact diagonalization. Our results lead to a band structure composed by a set of Landau levels, as given in Fig. \ref{Fig:LLs-Diagonalization}. The structure of the LL spectrum will be discussed in detail later.

The DOS of the system is calculated by using an
algorithm based on the evolution of the time-dependent Schr\"{o}dinger
equation. For this we use a random superposition of all basis states as
an initial state $|\varphi \rangle $
\begin{equation}
\left\vert \varphi \right\rangle =\sum_{i}a_{i}\left\vert i\right\rangle ,  \label{Eq:phi0}
\end{equation}%
where $a_{i}$ are random complex numbers normalized as $\sum_{i}\left\vert
a_{i}\right\vert ^{2}=1$, and the DOS is calculated as a Fourier transform of the
time-dependent correlation functions\cite{HR00,YRK10} 
\begin{equation}
d(\epsilon )=\frac{1}{2\pi }\int_{-\infty }^{\infty }e^{i\epsilon \tau
}\langle \varphi |e^{-iH\tau }|\varphi \rangle d\tau .  \label{Eq:DOS}
\end{equation}

\begin{figure*}[t]
\centering
 \includegraphics[width=1.0\textwidth]{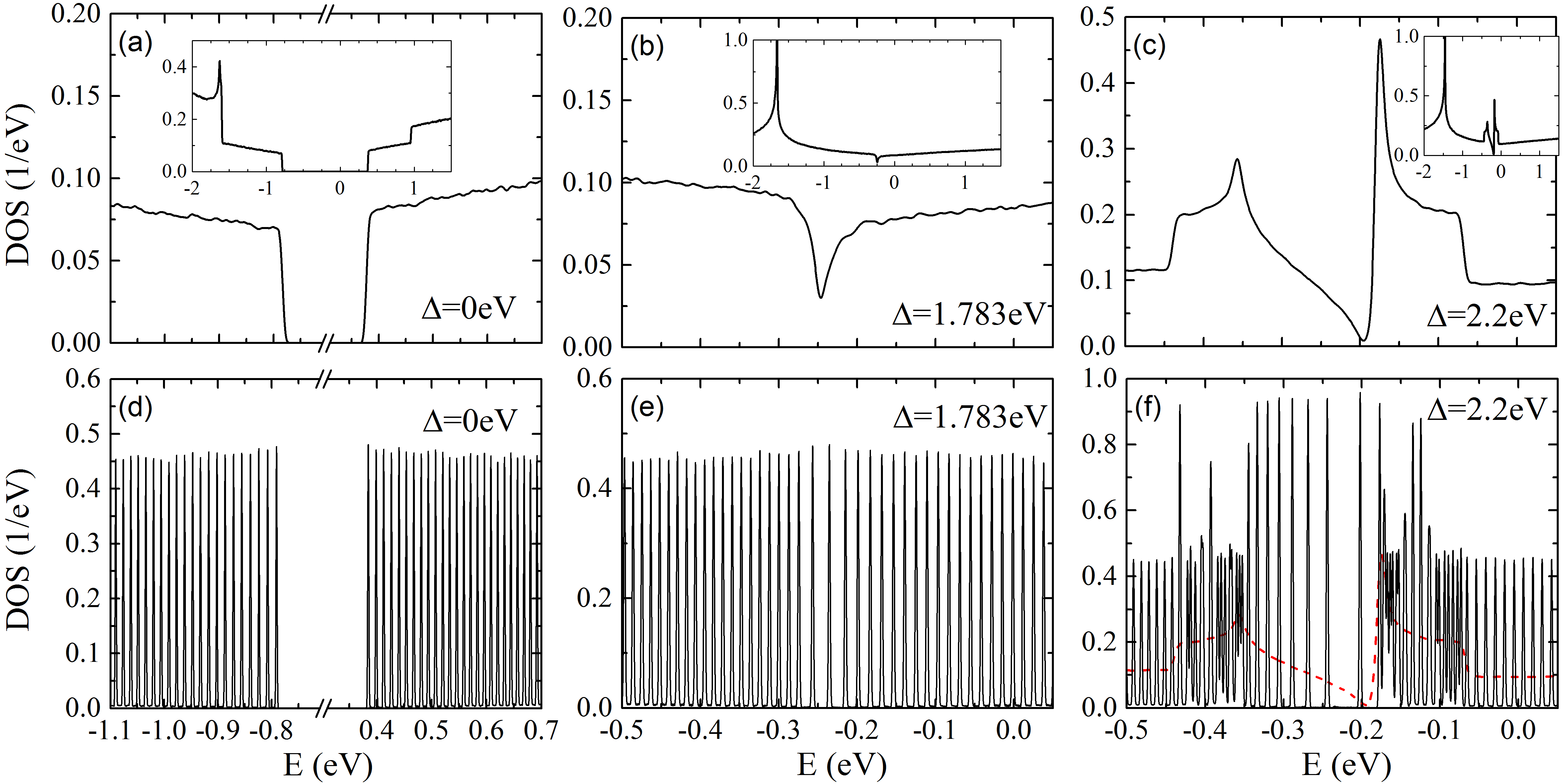}
\caption{Density of states and Landau level spectra of pristine and biased bilayer BP for the biased potential indicated in the figures. Panels (a)-(c) corresponds to $B=0$ and panels (d)-(f) to $B=65$~T. The calculated system contains $2\times 4800\times 4800$ atomic sites, with periodic boundary conditions on both $X$ and $Y$ directions. To illustrate the evolution with magnetic field of the DOS of the semimetalic phase ($\Delta=2.2$~eV), we include in panel (f) the DOS at $B=0$ (dashed red line). }
  \label{Fig:LLs}
\end{figure*}

In the unbiased semiconducting phase ($\Delta=0$) the DOS (per unit area) at low energies is approximately a constant $d_{\rm 2DEG}(\varepsilon)=gm_b/2\pi$ where $g=g_s=2$ is the spin degeneracy and $m^{c,v}_b=\sqrt{m^{c,v}_xm^{c,v}_y}$ is the band mass of the conduction ($c$) or valence ($v$) bands, as obtained in Fig. \ref{Fig:LLs}(a). In the presence of a quantifying magnetic field, the DOS is discretized into a set of Landau levels, as shown in Fig. \ref{Fig:LLs}(d) (see also Fig. \ref{Fig:LLs-Diagonalization}). We note that the finite broadening in the LLs is due to the energy resolution of the numerical simulations, which is limited
by the size of the sample used in the calculation (number of atoms), as well as the total number
of time steps, which determines the accuracy of the energy eigenvalues. The obtained LL spectrum consist of two sets of equidistant LLs separated by the band gap $E_g$ with energy $\varepsilon^{c,v}_n=\pm E_g/2\pm\omega^{c,v}_c(n+1/2)$ (where $n$ is a positive integer) separated by the cyclotron frequency $\omega_c^{c,v}=eB/m^{c,v}_b$. Since the system lacks of electron-hole symmetry, the cyclotron frequency is different for the valence and conduction bands. For $\Delta=\Delta_c=1.783~{\rm eV}$, the system suffers a semiconducting to semimetal transition, with a band crossing at the $\Gamma$ point [see Fig. \ref{Fig:Bands}(b)]. As it can be seen in Fig. \ref{Fig:LLs}(b), the DOS around such a band crossing is $\propto \sqrt{\varepsilon}$, leading to a set of non-equidistant LLs at energies close to the band crossing energy, with dispersion $\varepsilon_n\propto \pm [(n+1/2)B]^{2/3}$.\cite{DPM08,MG09b} As we move away from such band crossing, the LL spectrum has the same characteristics than the previous case of unbiased BP, recovering the standard quantization of a 2DEG [Fig. \ref{Fig:LLs}(e)].

The most interesting situation occurs for higher bias voltages, well beyond the transition. For $\Delta=2.2~{\rm eV}>\Delta_c$, the band dispersion present two Dirac points, in the $\Gamma-{\rm X}$ direction, and it is gapped in the $\Gamma-{\rm Y}$ direction [see Fig. \ref{Fig:Bands}(c) and \ref{Fig:CECs}(c)]. As studied by Montambaux {\it et al.} within the framework of an universal Hamiltonian that describes the merging of Dirac points in the electronic spectrum of 2D crystals,\cite{MG09,MG09b} the topological character of the transition can be understood from the appearance of a Berry phase for $\Delta>\Delta_c$, which takes the values $\pm \pi$ around each Dirac point. If we reduce the bias voltage, we recover the trivial phase with the corresponding annihilation of the Berry phase for $\Delta<\Delta_c$. For low carrier densities, the Fermi surface consists of two pockets encircling the Dirac points along the $\Gamma-{\rm X}$ direction [see Fig. \ref{Fig:CECs}(c)]. The DOS close to the Dirac points behaves as\cite{MG09b} 
$d_{\mathrm{Dirac}}(\epsilon)\propto|\epsilon|/{v_{Fx} v_{Fy}}$, where $v_{Fx(y)}$ is the Fermi velocity along the $x(y)$ direction within the Dirac cones. In a magnetic field, the LL spectrum is that of a semimetal with a {\it relativistic} quantization $\varepsilon_n\propto \pm \sqrt{nB}$ [see Fig. \ref{Fig:LLs}(f)]. The shift of $n+1/2\rightarrow n$ in the LL energy spectrum for $\Delta>\Delta_c$ is a consequence of the generation of $\pm \pi$ Berry phases around the Dirac points. If we increase the energy, we reach a highly nontrivial LL quantization because of the presence of a saddle point in the band structure, at which there is a transition from CECs encircling the Dirac points to CECs encircling the $\Gamma$ point.  In the semiclassical limit, the cyclotron orbits in reciprocal space follow the CECs. Therefore, at the saddle point there
is a change in the topological Berry phase from $\pm \pi $ for orbits encircling the Dirac points to 0 for orbits encircling the $\Gamma$ point.\cite{FM10,GGM12} The two series of LLs that we observe in Fig. \ref{Fig:LLs} is due to the different character of the cyclotron orbits at both sides of the
saddle point, with different cyclotron
frequencies, that \textit{merge} at the saddle point. This transition resembles that of highly doped graphene at energies around the Van Hove singularity.\cite{Hatsugai2006,YRK12}

\begin{figure*}[t]
\centering
 \includegraphics[width=0.32\textwidth]{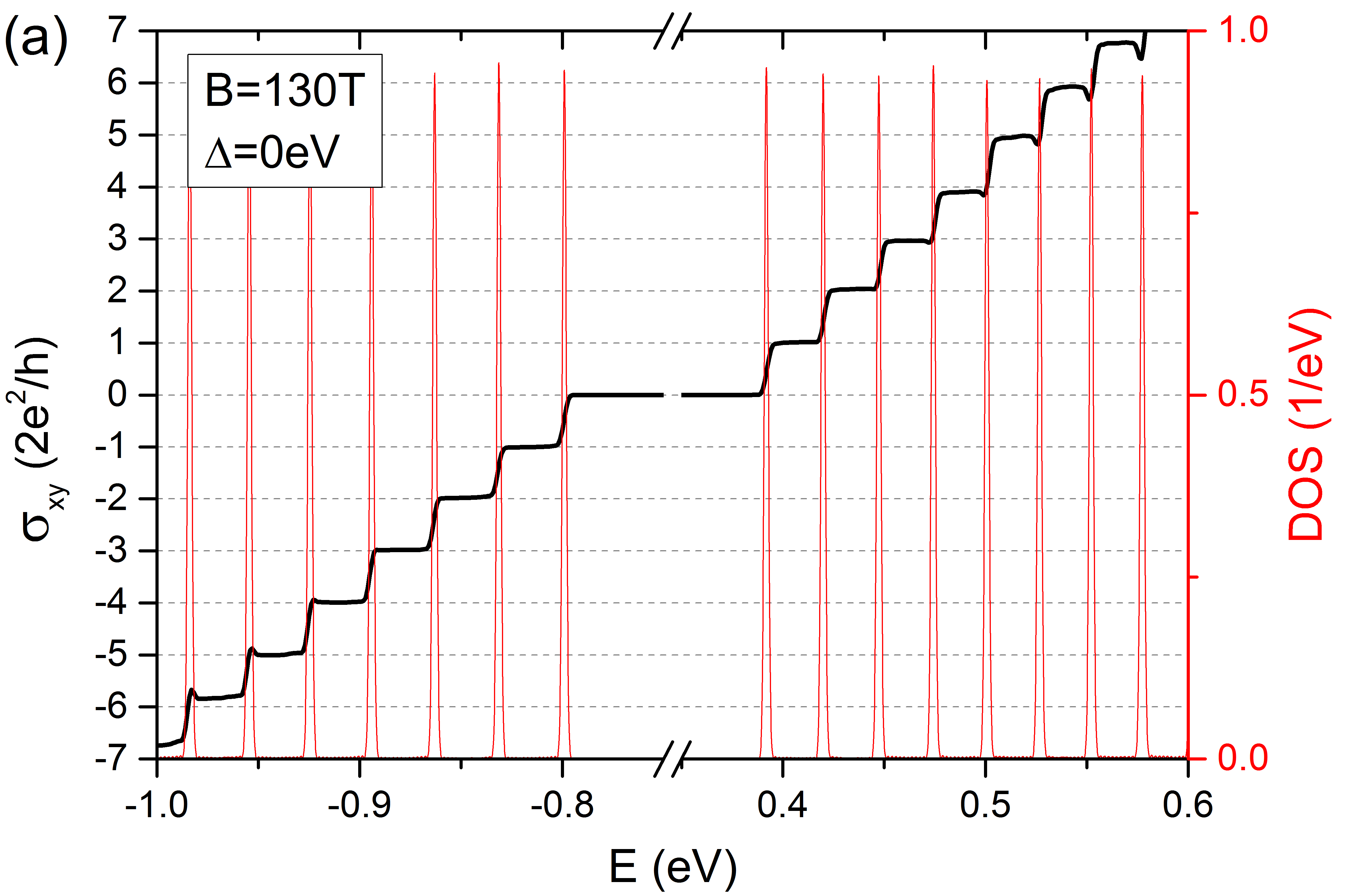}
 \includegraphics[width=0.32\textwidth]{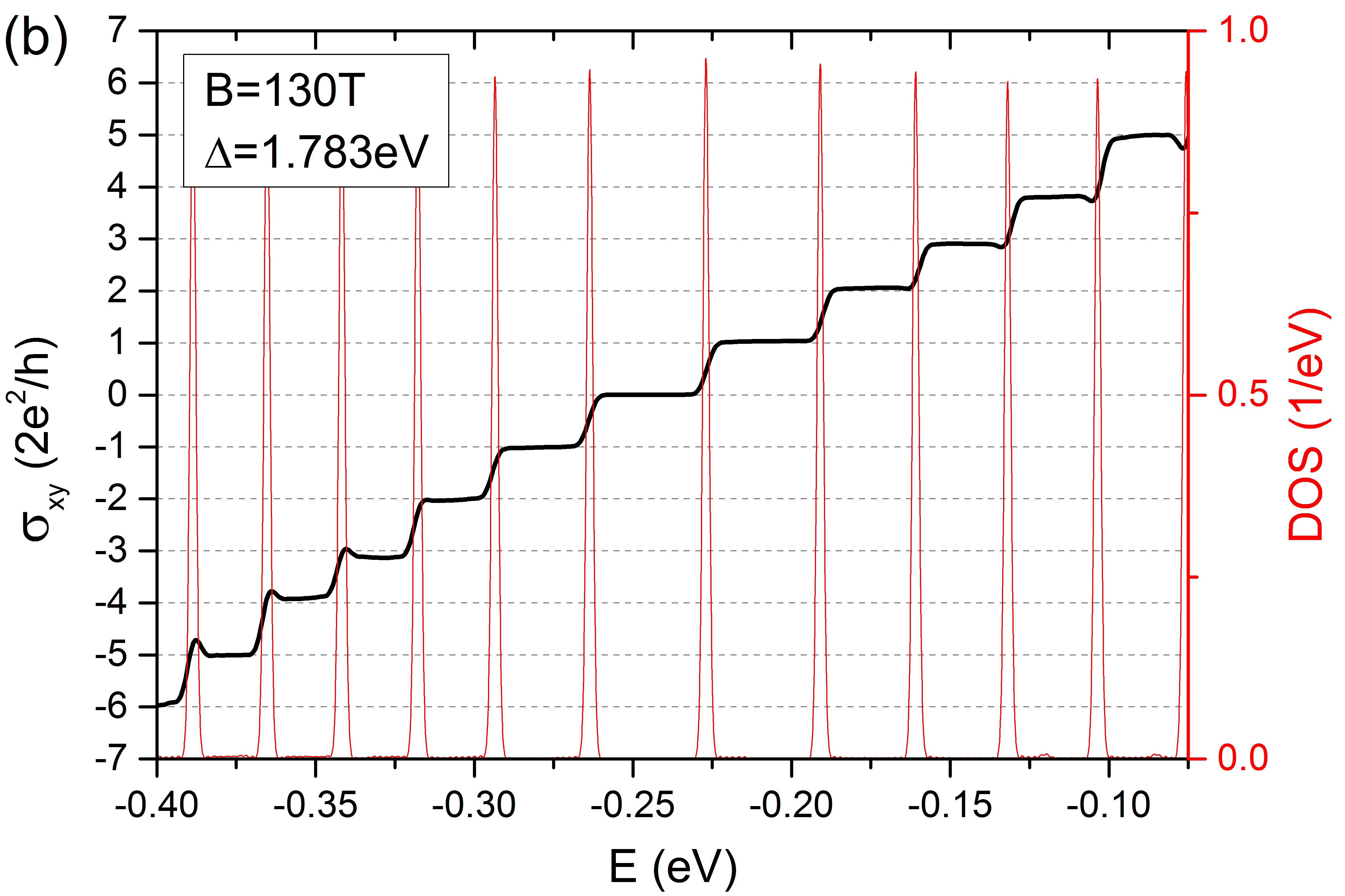}
  \includegraphics[width=0.32\textwidth]{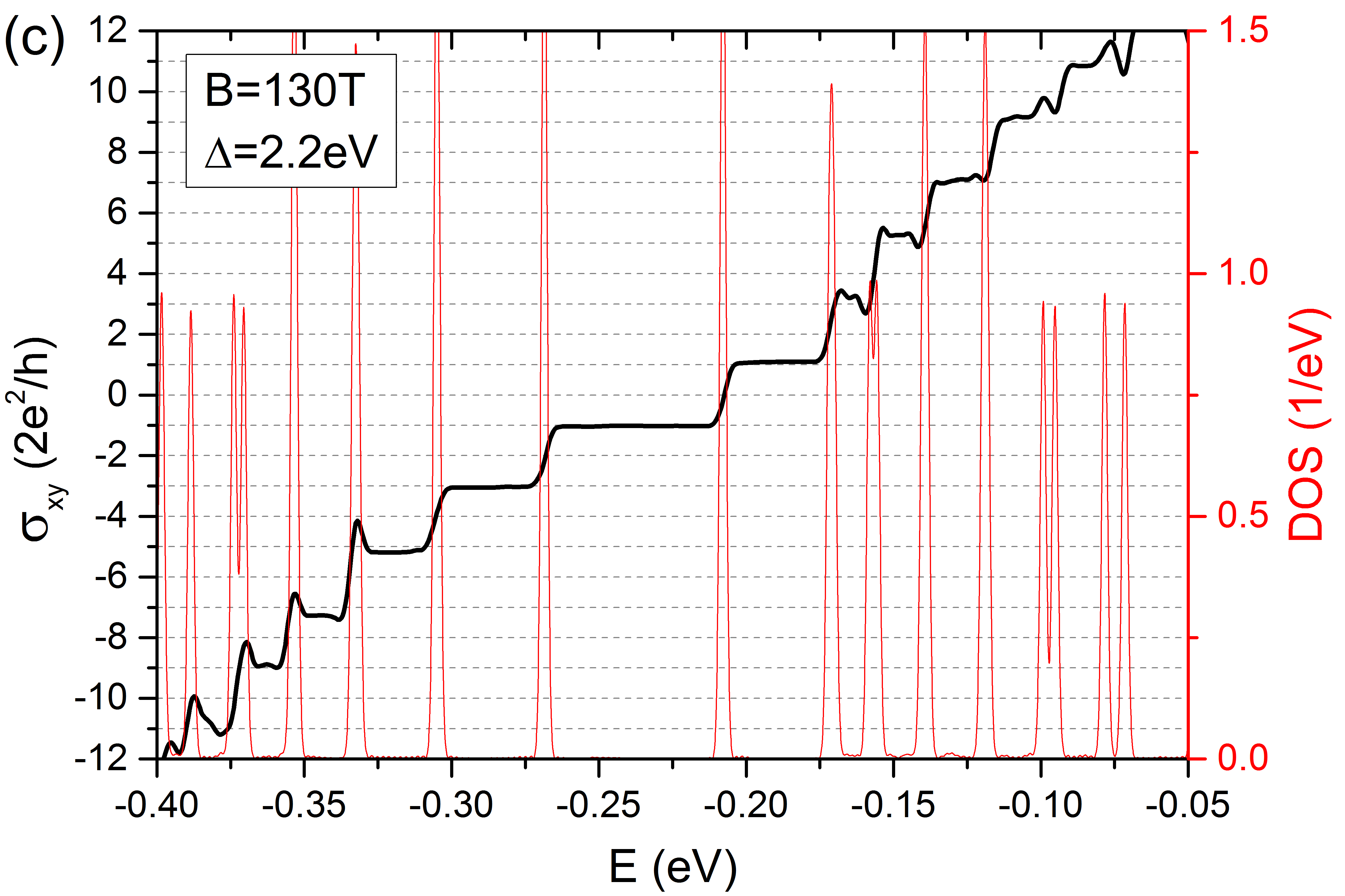}
\caption{Hall conductivity of pristine and biased BP for the biased potentials used in Fig. \ref{Fig:LLs}. The temperature is $T=0.01$~K and the magnetic field is $B=130$~T. The truncation order for the Kernel polynomial in Eq. (\ref{Eq:Kubo-Bastin-Numeric}) is $M=15000$. The calculated system contains $2\times 600\times 600$ atomic sites, and the results are averaged over five different random realization of initial states.}
  \label{Fig:Hall}
\end{figure*}

\section{Hall conductivity}

The next step in our analysis, once we understand the LL spectrum of the biased system, is the calculation  of the Hall conductivity $\sigma_{xy}$. For this aim, we use  an efficient numerical approach, recently developed by Garc\'{i}a {\it et al.},\cite{GCR15} that is based on a real space implementation of the Kubo formalism where both diagonal and off-diagonal conductivities are treated in the same footing.  In the limit $\omega\rightarrow 0$ and for non-interacting electrons, the so-called Kubo-Bastin formula for the conductivity can be used to obtain the elements of the static conductivity tensor\cite{Bastin71,GCR15,Ortmann2015}
\begin{eqnarray}\label{Eq:Kubo-Bastin}
\sigma_{{\alpha\beta}}(\mu,T)&=&\frac{i\hbar e^2}{A}\int_{-\infty}^{\infty}d\varepsilon f(\varepsilon){\rm Tr}\left \langle v_{\alpha}\delta(\varepsilon -{\cal H})v_{\beta}\frac{dG^+(\varepsilon)}{d\varepsilon}\right. \nonumber\\
&&\left.-v_{\alpha}\frac{dG^-(\varepsilon)}{d\varepsilon}v_{\beta}\delta(\varepsilon-{\cal H})\right\rangle,
\end{eqnarray}
where $\mu$ is the chemical potential, $T$ is the temperature, $A$ is the area of the sample, $v_{\alpha}$ is the $\alpha$ component of the velocity operator, $G^{\pm}(\varepsilon)=1/(\varepsilon-{\cal H}\pm i\eta)$ are the Green's functions, and $f(\varepsilon)$ is the Fermi-Dirac distribution. Here the average is performed by using the same random initial state as in the calculation of DOS. By expanding the delta and the Green's functions $G^{\pm}(\varepsilon)$ in terms of Chebyshev polynomials (using the so-called kernel polynomial method),\cite{GCR15} the conductivity tensor becomes
\begin{eqnarray}
\sigma _{{\alpha \beta }}(\mu ,T) &=&\frac{4e^{2}\hbar }{\pi A}\frac{4}{%
\Delta E^{2}}\int_{-1}^{1}d\tilde{\varepsilon}\frac{f(\tilde{\varepsilon})}{%
(1-\tilde{\varepsilon}^{2})^{2}}\sum_{m,n}\Gamma _{{nm}}(\tilde{\varepsilon}%
)\mu _{{nm}}^{\alpha \beta }(\tilde{H}),  \notag  \label{Eq:Kubo-Bastin-Numeric} \\
&&
\end{eqnarray}
where $\Delta E$ is the energy range of the spectrum, $\tilde { \varepsilon }$ is the rescaled energy within [-1,1], $\Gamma_{mn}(\tilde{\varepsilon})$ and $\mu^{\alpha\beta}_{{mn}}(\tilde{H})$ are functions of the energy and the Hamiltonian, respectively. More precisely,
\begin{eqnarray}
\Gamma_{mn}(\tilde{\varepsilon})&\equiv &T_m(\tilde{\varepsilon})(\tilde{\varepsilon}-in\sqrt{1-\tilde{\varepsilon}^2})e^{in\arccos(\tilde{\varepsilon})}\nonumber\\
&+&T_n(\tilde{\varepsilon})(\tilde{\varepsilon}+im\sqrt{1-\tilde{\varepsilon}^2})e^{-im\arccos(\tilde{\varepsilon})}
\end{eqnarray}
is a scalar function of the rescaled energy, and
 \begin{equation}
\mu^{\alpha\beta}_{{mn}}(\tilde{H})\equiv \frac{g_m g_n}{(1+\delta_{{n0}})(1+\delta_{{m0}})}\text{Tr}\left[v_\alpha T_m(\tilde{H})v_\beta
T_n(\tilde{H})\right]
\end{equation}
 is independent of the energy, where $T_m(x)$ are the Chebyshev polynomial defined according to the recurrence relation $T_{m}(x)=2xT_{m-1}(x)-T_{m-2}(x)$ with $T_{0}(x)=1$ and $T_{1}(x)=x$. The Gibbs oscillations due to the truncation of the expansion in  (\ref{Eq:Kubo-Bastin-Numeric}) are smoothed by using the Jackson Kernel $g_m$.~\cite{KPM,GCR15}

Our results for $\sigma_{xy}$ are shown in Fig. \ref{Fig:Hall}. For $\Delta=0$ the Hall conductivity consist on a series of plateaus with the well known sequence $\sigma_{xy}=2ne^2/h$, characteristics of a standard 2DEG with a parabolic band dispersion (although  the present case of BP is rather described by a paraboloidal band). Our numerical calculations show the same quantization of the Hall conductivity at the transition point $\Delta=\Delta_c$ [Fig. \ref{Fig:Hall}c]. This is due to the fact that, right at the transition, there is a single crossing of the bands at the time-reversal invariant $\Gamma$ point of the Brillouin zone.\cite{MG09,MG09b}  Most saliently, for $\Delta>\Delta_c$ (Fig.\ref{Fig:Hall}c) the Hall conductivity presents plateaus at $\sigma_{xy}=4(n+1/2)e^2/h$. This is due to the topological nature of the semimetalic phase, which is well captured by the numerical method. The plateau structure becomes blurred at high energies, which is an artifact due to finite truncation order of the Kernel polynomial approximation as well as the finite size of the sample.\cite{GCR15} These artifacts can be improved with a higher truncation order of the expansion, and by considering a larger sample size. This would lead for an initial state, obtained from Eq. (\ref{Eq:phi0}), to be a more accurate representation of the whole energy spectrum.\cite{HR00,YRK10} Furthermore it has been shown that the convergence of the Hall conductivity in Eq. (\ref{Eq:Kubo-Bastin-Numeric}) is faster with larger magnetic fields.\cite{GCR15} Therefore, in order to catch several Landau levels within the emerged Dirac cones, a large magnetic field of $B=130$~T is used in the calculations. Lower magnetic fields will give similar qualitative behavior for the Landau level spectrum and for the Hall conductivity. In spite of the above choices in the simulation, the convergence of the results is still slow, especially if the band structure contains different topological features within a small energy range, such as in the case with $\Delta>\Delta_c$. We adopted a truncation order as large as $M=15000$ for the Kernel polynomial decomposition (the computational costs are proportional to $M^2$ and the maximum truncation order used in Ref. \onlinecite{GCR15} is $M=6144$), and the simulated systems consist of $2\times 600\times 600$ atomic sites. However, it is still not enough to overcome the blurred effects in the high energy plateaus. Further calculations with larger truncation order or sample size are beyond the computational resource that we can reach.

\section{Discussion and conclusions}

We notice that the emergence of Dirac points in the spectrum of biased black phosphorus can be understood by thinking of the BP lattice as a honeycomb lattice (like the one in graphene) in which one of the three hopping terms between nearest-neighbour atoms can be different from the other two.\cite{MG09b} This is indeed the case in BP, in which two of the three nearest neighbors of one atom are in the same plane, whereas the third nearest neighbor is in a different plane. Moreover, the signs of those hopping terms are different, making BP a natural platform to realize Dirac points engineering near the $\Gamma$ point,\cite{MG09} either by tuning external bias or by applying  strain  to the samples.

In summary, we have analyzed the electronic properties of biased black phosphorus in the presence of a perpendicular magnetic field. In the absence of an electric field, the external magnetic field leads to a quantization of the electron and hole bands into set of equidistant Landau levels. This behavior is similar to the discretization of the energy dispersion in a 2DEG with a parabolic band.  If we further apply a perpendicular electric field to the sample, we obtain a reduction of the band gap with the applied voltage. For a critical value of the voltage, the gap completely closes, and a pair of Dirac points appear in the $\Gamma-{\rm X}$ direction of the Brillouin zone. This semiconductor to semimetal transition is accompanied  by a change in the topology of the system, due to the generation of $\pm \pi$ Berry phases around the Dirac points. We obtain a highly nontrivial Landau level spectrum in this phase, with a coexistence of relativistic Landau levels, with a $\varepsilon_n\propto \sqrt{nB}$ quantization, with equidistant LLs at higher energies, following the standard $\varepsilon_n\propto B(n+1/2)$, characteristic of a 2DEG. The transition between these two regimes requires to go through a Van Hove singularity (saddle point) in the band dispersion, with the corresponding divergence in the density of states. Finally, we numerically compute the Hall conductivity of the system. The topological transition driven by the electric field is reflected in a different quantization of the Hall conductivity, which present the characteristic $\sigma_{xy}\propto 2n$ behavior for small bias voltages (insufficient to close the gap), and a {\it relativistic} quantum Hall effect with $\sigma_{xy}\propto 4(n+1/2)$ in the semimetalic phase, due to the generation of a pair of Dirac cones. Although we focus on the simplest case of bilayer BP, the results presented here apply to any multilayer BP sample, with the advantage that the gap decreases with the number of layers, and therefore the semiconducting to semimetal transition would be more easily reached for thicker samples. We notice that the electric field induced semimetallic phase in BP is likely to present new broken symmetry phases due to many body effects, which are not included here. For example, it is known that bilayer graphene, whose low energy spectrum reassembles that of biased semimetalic BP, suffers a nematic phase transition driven by Coulomb interactions.\cite{Mayorov_Science2011} Similar interaction-driven phase transitions might occur in BP and will be the object of future studies. The phenomena discussed here could be observed by exposing to a strong quantizing magnetic field a biased BP sample, chemically doped from in-situ deposition of adatoms,\cite{KK15} or by applying external strain (compression) to the samples.\cite{XC15} These techniques have been shown to be appropriate routes to tune this material from a moderate-gap semiconductor to a band-inverted semimetal.

\acknowledgments

We are grateful to G. Montambaux, J.-N. F\"uchs and M. O. Goerbig for very useful discussions. The support by the Stichting Fundamenteel Onderzoek der Materie (FOM) and
the Netherlands National Computing Facilities foundation (NCF) are
acknowledged. S.Y. and M.I.K. thank financial support from the European
Research Council Advanced Grant program (contract 338957). 
This project has received funding from the European UnionÕs Horizon  2020 research and innovation programme under grant agreement No. 696656 Ð GrapheneCore1. 
R.R. acknowledges financial support from MINECO (Spain) through grant FIS2014-58445-JIN, and project No. PIB2010BZ-00512, from the Comunidad Aut\'onoma de Madrid
(CAM) MAD2D-CM Program (S2013/MIT-3007).

\bibliography{bp,Bib_2D}

\newcommand{\npb}{Nucl. Phys.}\newcommand{\adv}{Adv.
  Phys.}\newcommand{\epl}{Europhys. Lett.}
\begin{thebibliography}{48}
\expandafter\ifx\csname natexlab\endcsname\relax\def\natexlab#1{#1}\fi
\expandafter\ifx\csname bibnamefont\endcsname\relax
  \def\bibnamefont#1{#1}\fi
\expandafter\ifx\csname bibfnamefont\endcsname\relax
  \def\bibfnamefont#1{#1}\fi
\expandafter\ifx\csname citenamefont\endcsname\relax
  \def\citenamefont#1{#1}\fi
\expandafter\ifx\csname url\endcsname\relax
  \def\url#1{\texttt{#1}}\fi
\expandafter\ifx\csname urlprefix\endcsname\relax\def\urlprefix{URL }\fi
\providecommand{\bibinfo}[2]{#2}
\providecommand{\eprint}[2][]{\url{#2}}

\bibitem[{\citenamefont{Li et~al.}(2014)\citenamefont{Li, Yu, Ye, Ge, Ou, Wu,
  Feng, Chen, and Zhang}}]{li14}
\bibinfo{author}{\bibfnamefont{L.}~\bibnamefont{Li}},
  \bibinfo{author}{\bibfnamefont{Y.}~\bibnamefont{Yu}},
  \bibinfo{author}{\bibfnamefont{G.~J.} \bibnamefont{Ye}},
  \bibinfo{author}{\bibfnamefont{Q.}~\bibnamefont{Ge}},
  \bibinfo{author}{\bibfnamefont{X.}~\bibnamefont{Ou}},
  \bibinfo{author}{\bibfnamefont{H.}~\bibnamefont{Wu}},
  \bibinfo{author}{\bibfnamefont{D.}~\bibnamefont{Feng}},
  \bibinfo{author}{\bibfnamefont{X.~H.} \bibnamefont{Chen}}, \bibnamefont{and}
  \bibinfo{author}{\bibfnamefont{Y.}~\bibnamefont{Zhang}},
  \bibinfo{journal}{Nature Nanotechnology} \textbf{\bibinfo{volume}{9}},
  \bibinfo{pages}{372} (\bibinfo{year}{2014}).

\bibitem[{\citenamefont{Xia et~al.}(2014)\citenamefont{Xia, Wang, and
  Jia}}]{xia1}
\bibinfo{author}{\bibfnamefont{F.}~\bibnamefont{Xia}},
  \bibinfo{author}{\bibfnamefont{H.}~\bibnamefont{Wang}}, \bibnamefont{and}
  \bibinfo{author}{\bibfnamefont{Y.}~\bibnamefont{Jia}},
  \bibinfo{journal}{Nature Communications} \textbf{\bibinfo{volume}{5}}
  (\bibinfo{year}{2014}).

\bibitem[{\citenamefont{Castellanos-Gomez}(2015)}]{C15}
\bibinfo{author}{\bibfnamefont{A.}~\bibnamefont{Castellanos-Gomez}},
  \bibinfo{journal}{The Journal of Physical Chemistry Letters}
  \textbf{\bibinfo{volume}{6}}, \bibinfo{pages}{4280} (\bibinfo{year}{2015}).

\bibitem[{\citenamefont{Morita}(1986)}]{morita86}
\bibinfo{author}{\bibfnamefont{A.}~\bibnamefont{Morita}},
  \bibinfo{journal}{Applied Physics A} \textbf{\bibinfo{volume}{39}},
  \bibinfo{pages}{227} (\bibinfo{year}{1986}).

\bibitem[{\citenamefont{Liu et~al.}(2014)\citenamefont{Liu, Neal, Zhu, Luo, Xu,
  Tom{\'a}nek, and Ye}}]{liu14}
\bibinfo{author}{\bibfnamefont{H.}~\bibnamefont{Liu}},
  \bibinfo{author}{\bibfnamefont{A.~T.} \bibnamefont{Neal}},
  \bibinfo{author}{\bibfnamefont{Z.}~\bibnamefont{Zhu}},
  \bibinfo{author}{\bibfnamefont{Z.}~\bibnamefont{Luo}},
  \bibinfo{author}{\bibfnamefont{X.}~\bibnamefont{Xu}},
  \bibinfo{author}{\bibfnamefont{D.}~\bibnamefont{Tom{\'a}nek}},
  \bibnamefont{and} \bibinfo{author}{\bibfnamefont{P.~D.} \bibnamefont{Ye}},
  \bibinfo{journal}{ACS nano} \textbf{\bibinfo{volume}{8}},
  \bibinfo{pages}{4033} (\bibinfo{year}{2014}).

\bibitem[{\citenamefont{Qiao et~al.}(2014)\citenamefont{Qiao, Kong, Hu, Yang,
  and Ji}}]{qiao14}
\bibinfo{author}{\bibfnamefont{J.}~\bibnamefont{Qiao}},
  \bibinfo{author}{\bibfnamefont{X.}~\bibnamefont{Kong}},
  \bibinfo{author}{\bibfnamefont{Z.-X.} \bibnamefont{Hu}},
  \bibinfo{author}{\bibfnamefont{F.}~\bibnamefont{Yang}}, \bibnamefont{and}
  \bibinfo{author}{\bibfnamefont{W.}~\bibnamefont{Ji}},
  \bibinfo{journal}{Nature Communications} \textbf{\bibinfo{volume}{5}}
  (\bibinfo{year}{2014}).

\bibitem[{\citenamefont{Low et~al.}(2014{\natexlab{a}})\citenamefont{Low,
  Rodin, Carvalho, Jiang, Wang, Xia, and Neto}}]{low14cond}
\bibinfo{author}{\bibfnamefont{T.}~\bibnamefont{Low}},
  \bibinfo{author}{\bibfnamefont{A.}~\bibnamefont{Rodin}},
  \bibinfo{author}{\bibfnamefont{A.}~\bibnamefont{Carvalho}},
  \bibinfo{author}{\bibfnamefont{Y.}~\bibnamefont{Jiang}},
  \bibinfo{author}{\bibfnamefont{H.}~\bibnamefont{Wang}},
  \bibinfo{author}{\bibfnamefont{F.}~\bibnamefont{Xia}}, \bibnamefont{and}
  \bibinfo{author}{\bibfnamefont{A.~C.} \bibnamefont{Neto}},
  \bibinfo{journal}{Physical Review B} \textbf{\bibinfo{volume}{90}},
  \bibinfo{pages}{075434} (\bibinfo{year}{2014}{\natexlab{a}}).

\bibitem[{\citenamefont{Yuan et~al.}(2015)\citenamefont{Yuan, Rudenko, and
  Katsnelson}}]{YRK15}
\bibinfo{author}{\bibfnamefont{S.}~\bibnamefont{Yuan}},
  \bibinfo{author}{\bibfnamefont{A.~N.} \bibnamefont{Rudenko}},
  \bibnamefont{and} \bibinfo{author}{\bibfnamefont{M.~I.}
  \bibnamefont{Katsnelson}}, \bibinfo{journal}{Phys. Rev. B}
  \textbf{\bibinfo{volume}{91}}, \bibinfo{pages}{115436}
  (\bibinfo{year}{2015}).

\bibitem[{\citenamefont{Low et~al.}(2014{\natexlab{b}})\citenamefont{Low,
  Rold{\'a}n, Wang, Xia, Avouris, Moreno, and Guinea}}]{low14plas}
\bibinfo{author}{\bibfnamefont{T.}~\bibnamefont{Low}},
  \bibinfo{author}{\bibfnamefont{R.}~\bibnamefont{Rold{\'a}n}},
  \bibinfo{author}{\bibfnamefont{H.}~\bibnamefont{Wang}},
  \bibinfo{author}{\bibfnamefont{F.}~\bibnamefont{Xia}},
  \bibinfo{author}{\bibfnamefont{P.}~\bibnamefont{Avouris}},
  \bibinfo{author}{\bibfnamefont{L.~M.} \bibnamefont{Moreno}},
  \bibnamefont{and} \bibinfo{author}{\bibfnamefont{F.}~\bibnamefont{Guinea}},
  \bibinfo{journal}{Physical Review Letters} \textbf{\bibinfo{volume}{113}},
  \bibinfo{pages}{106802} (\bibinfo{year}{2014}{\natexlab{b}}).

\bibitem[{\citenamefont{Chen et~al.}(2014)\citenamefont{Chen, Wu, Wu, Xu, Wang,
  Han, Ye, Han, He, Cai et~al.}}]{exp1}
\bibinfo{author}{\bibfnamefont{X.}~\bibnamefont{Chen}},
  \bibinfo{author}{\bibfnamefont{Y.}~\bibnamefont{Wu}},
  \bibinfo{author}{\bibfnamefont{Z.}~\bibnamefont{Wu}},
  \bibinfo{author}{\bibfnamefont{S.}~\bibnamefont{Xu}},
  \bibinfo{author}{\bibfnamefont{L.}~\bibnamefont{Wang}},
  \bibinfo{author}{\bibfnamefont{Y.}~\bibnamefont{Han}},
  \bibinfo{author}{\bibfnamefont{W.}~\bibnamefont{Ye}},
  \bibinfo{author}{\bibfnamefont{T.}~\bibnamefont{Han}},
  \bibinfo{author}{\bibfnamefont{Y.}~\bibnamefont{He}},
  \bibinfo{author}{\bibfnamefont{Y.}~\bibnamefont{Cai}}, \bibnamefont{et~al.},
  \bibinfo{journal}{Nature Communications} \textbf{\bibinfo{volume}{6}},
  \bibinfo{pages}{7315} (\bibinfo{year}{2014}).

\bibitem[{\citenamefont{Gillgren et~al.}(2015)\citenamefont{Gillgren,
  Wickramaratne, Shi, Espiritu, Yang, Hu, Wei, Liu, Mao, Watanabe
  et~al.}}]{exp2}
\bibinfo{author}{\bibfnamefont{N.}~\bibnamefont{Gillgren}},
  \bibinfo{author}{\bibfnamefont{D.}~\bibnamefont{Wickramaratne}},
  \bibinfo{author}{\bibfnamefont{Y.}~\bibnamefont{Shi}},
  \bibinfo{author}{\bibfnamefont{T.}~\bibnamefont{Espiritu}},
  \bibinfo{author}{\bibfnamefont{J.}~\bibnamefont{Yang}},
  \bibinfo{author}{\bibfnamefont{J.}~\bibnamefont{Hu}},
  \bibinfo{author}{\bibfnamefont{J.}~\bibnamefont{Wei}},
  \bibinfo{author}{\bibfnamefont{X.}~\bibnamefont{Liu}},
  \bibinfo{author}{\bibfnamefont{Z.}~\bibnamefont{Mao}},
  \bibinfo{author}{\bibfnamefont{K.}~\bibnamefont{Watanabe}},
  \bibnamefont{et~al.}, \bibinfo{journal}{2D Materials}
  \textbf{\bibinfo{volume}{2}}, \bibinfo{pages}{011001} (\bibinfo{year}{2015}).

\bibitem[{\citenamefont{Li et~al.}(2015)\citenamefont{Li, Ye, Tran, Fei, Chen,
  Wang, Wang, Watanabe, Taniguchi, Yang et~al.}}]{exp3}
\bibinfo{author}{\bibfnamefont{L.}~\bibnamefont{Li}},
  \bibinfo{author}{\bibfnamefont{G.~J.} \bibnamefont{Ye}},
  \bibinfo{author}{\bibfnamefont{V.}~\bibnamefont{Tran}},
  \bibinfo{author}{\bibfnamefont{R.}~\bibnamefont{Fei}},
  \bibinfo{author}{\bibfnamefont{G.}~\bibnamefont{Chen}},
  \bibinfo{author}{\bibfnamefont{H.}~\bibnamefont{Wang}},
  \bibinfo{author}{\bibfnamefont{J.}~\bibnamefont{Wang}},
  \bibinfo{author}{\bibfnamefont{K.}~\bibnamefont{Watanabe}},
  \bibinfo{author}{\bibfnamefont{T.}~\bibnamefont{Taniguchi}},
  \bibinfo{author}{\bibfnamefont{L.}~\bibnamefont{Yang}}, \bibnamefont{et~al.},
  \bibinfo{journal}{Nature Nanotechnology} \textbf{\bibinfo{volume}{10}},
  \bibinfo{pages}{608} (\bibinfo{year}{2015}).

\bibitem[{\citenamefont{Tayari et~al.}(2014)\citenamefont{Tayari, Hemsworth,
  Fakih, Favron, Gaufr{\`e}s, Gervais, Martel, and Szkopek}}]{exp4}
\bibinfo{author}{\bibfnamefont{V.}~\bibnamefont{Tayari}},
  \bibinfo{author}{\bibfnamefont{N.}~\bibnamefont{Hemsworth}},
  \bibinfo{author}{\bibfnamefont{I.}~\bibnamefont{Fakih}},
  \bibinfo{author}{\bibfnamefont{A.}~\bibnamefont{Favron}},
  \bibinfo{author}{\bibfnamefont{E.}~\bibnamefont{Gaufr{\`e}s}},
  \bibinfo{author}{\bibfnamefont{G.}~\bibnamefont{Gervais}},
  \bibinfo{author}{\bibfnamefont{R.}~\bibnamefont{Martel}}, \bibnamefont{and}
  \bibinfo{author}{\bibfnamefont{T.}~\bibnamefont{Szkopek}},
  \bibinfo{journal}{Nature Communications} \textbf{\bibinfo{volume}{6}},
  \bibinfo{pages}{7702} (\bibinfo{year}{2014}).

\bibitem[{\citenamefont{Cao et~al.}(2015)\citenamefont{Cao, Mishchenko, Yu,
  Khestanova, Rooney, Prestat, Kretinin, Blake, Shalom, Balakrishnan
  et~al.}}]{cao15}
\bibinfo{author}{\bibfnamefont{Y.}~\bibnamefont{Cao}},
  \bibinfo{author}{\bibfnamefont{A.}~\bibnamefont{Mishchenko}},
  \bibinfo{author}{\bibfnamefont{G.}~\bibnamefont{Yu}},
  \bibinfo{author}{\bibfnamefont{K.}~\bibnamefont{Khestanova}},
  \bibinfo{author}{\bibfnamefont{A.}~\bibnamefont{Rooney}},
  \bibinfo{author}{\bibfnamefont{E.}~\bibnamefont{Prestat}},
  \bibinfo{author}{\bibfnamefont{A.}~\bibnamefont{Kretinin}},
  \bibinfo{author}{\bibfnamefont{P.}~\bibnamefont{Blake}},
  \bibinfo{author}{\bibfnamefont{M.}~\bibnamefont{Shalom}},
  \bibinfo{author}{\bibfnamefont{G.}~\bibnamefont{Balakrishnan}},
  \bibnamefont{et~al.}, \bibinfo{journal}{Nano Letters}
  \textbf{\bibinfo{volume}{15}}, \bibinfo{pages}{4914} (\bibinfo{year}{2015}).

\bibitem[{\citenamefont{{Li} et~al.}(2015)\citenamefont{{Li}, {Yang}, {Ye},
  {Zhang}, {Zhu}, {Lou}, {Li}, {Watanabe}, {Taniguchi}, {Chang}
  et~al.}}]{Li2015}
\bibinfo{author}{\bibfnamefont{L.}~\bibnamefont{{Li}}},
  \bibinfo{author}{\bibfnamefont{F.}~\bibnamefont{{Yang}}},
  \bibinfo{author}{\bibfnamefont{G.~J.} \bibnamefont{{Ye}}},
  \bibinfo{author}{\bibfnamefont{Z.}~\bibnamefont{{Zhang}}},
  \bibinfo{author}{\bibfnamefont{Z.}~\bibnamefont{{Zhu}}},
  \bibinfo{author}{\bibfnamefont{W.-K.} \bibnamefont{{Lou}}},
  \bibinfo{author}{\bibfnamefont{L.}~\bibnamefont{{Li}}},
  \bibinfo{author}{\bibfnamefont{K.}~\bibnamefont{{Watanabe}}},
  \bibinfo{author}{\bibfnamefont{T.}~\bibnamefont{{Taniguchi}}},
  \bibinfo{author}{\bibfnamefont{K.}~\bibnamefont{{Chang}}},
  \bibnamefont{et~al.}, \bibinfo{journal}{ArXiv e-prints}
  (\bibinfo{year}{2015}), \eprint{1504.07155}.

\bibitem[{\citenamefont{Kim et~al.}(2015)\citenamefont{Kim, Baik, Ryu, Sohn,
  Park, Park, Denlinger, Yi, Choi, and Kim}}]{KK15}
\bibinfo{author}{\bibfnamefont{J.}~\bibnamefont{Kim}},
  \bibinfo{author}{\bibfnamefont{S.~S.} \bibnamefont{Baik}},
  \bibinfo{author}{\bibfnamefont{S.~H.} \bibnamefont{Ryu}},
  \bibinfo{author}{\bibfnamefont{Y.}~\bibnamefont{Sohn}},
  \bibinfo{author}{\bibfnamefont{S.}~\bibnamefont{Park}},
  \bibinfo{author}{\bibfnamefont{B.-G.} \bibnamefont{Park}},
  \bibinfo{author}{\bibfnamefont{J.}~\bibnamefont{Denlinger}},
  \bibinfo{author}{\bibfnamefont{Y.}~\bibnamefont{Yi}},
  \bibinfo{author}{\bibfnamefont{H.~J.} \bibnamefont{Choi}}, \bibnamefont{and}
  \bibinfo{author}{\bibfnamefont{K.~S.} \bibnamefont{Kim}},
  \bibinfo{journal}{Science} \textbf{\bibinfo{volume}{349}},
  \bibinfo{pages}{723} (\bibinfo{year}{2015}).

\bibitem[{\citenamefont{Liu et~al.}(2015)\citenamefont{Liu, Zhang, Abdalla,
  Fazzio, and Zunger}}]{LZ15}
\bibinfo{author}{\bibfnamefont{Q.}~\bibnamefont{Liu}},
  \bibinfo{author}{\bibfnamefont{X.}~\bibnamefont{Zhang}},
  \bibinfo{author}{\bibfnamefont{L.~B.} \bibnamefont{Abdalla}},
  \bibinfo{author}{\bibfnamefont{A.}~\bibnamefont{Fazzio}}, \bibnamefont{and}
  \bibinfo{author}{\bibfnamefont{A.}~\bibnamefont{Zunger}},
  \bibinfo{journal}{Nano Letters} \textbf{\bibinfo{volume}{15}},
  \bibinfo{pages}{1222} (\bibinfo{year}{2015}).

\bibitem[{\citenamefont{{Dolui} and {Quek}}(2015)}]{DQ15}
\bibinfo{author}{\bibfnamefont{K.}~\bibnamefont{{Dolui}}} \bibnamefont{and}
  \bibinfo{author}{\bibfnamefont{S.~Y.} \bibnamefont{{Quek}}},
  \bibinfo{journal}{Scientific Reports} \textbf{\bibinfo{volume}{5}},
  \bibinfo{pages}{11699} (\bibinfo{year}{2015}).

\bibitem[{\citenamefont{Chaves et~al.}(2015)\citenamefont{Chaves, Low, Avouris,
  \ifmmode \mbox{\c{C}}\else \c{C}\fi{}ak\ifmmode \imath \else~\i \fi{}r, and
  Peeters}}]{Chaves2015}
\bibinfo{author}{\bibfnamefont{A.}~\bibnamefont{Chaves}},
  \bibinfo{author}{\bibfnamefont{T.}~\bibnamefont{Low}},
  \bibinfo{author}{\bibfnamefont{P.}~\bibnamefont{Avouris}},
  \bibinfo{author}{\bibfnamefont{D.}~\bibnamefont{\ifmmode \mbox{\c{C}}\else
  \c{C}\fi{}ak\ifmmode \imath \else~\i \fi{}r}}, \bibnamefont{and}
  \bibinfo{author}{\bibfnamefont{F.~M.} \bibnamefont{Peeters}},
  \bibinfo{journal}{Phys. Rev. B} \textbf{\bibinfo{volume}{91}},
  \bibinfo{pages}{155311} (\bibinfo{year}{2015}).

\bibitem[{\citenamefont{Baik et~al.}(2015)\citenamefont{Baik, Kim, Yi, and
  Choi}}]{Baik_2015}
\bibinfo{author}{\bibfnamefont{S.~S.} \bibnamefont{Baik}},
  \bibinfo{author}{\bibfnamefont{K.~S.} \bibnamefont{Kim}},
  \bibinfo{author}{\bibfnamefont{Y.}~\bibnamefont{Yi}}, \bibnamefont{and}
  \bibinfo{author}{\bibfnamefont{H.~J.} \bibnamefont{Choi}},
  \bibinfo{journal}{Nano Letters} \textbf{\bibinfo{volume}{15}},
  \bibinfo{pages}{7788} (\bibinfo{year}{2015}).

\bibitem[{\citenamefont{Rodin et~al.}(2014)\citenamefont{Rodin, Carvalho, and
  Castro~Neto}}]{Rodin2014}
\bibinfo{author}{\bibfnamefont{A.~S.} \bibnamefont{Rodin}},
  \bibinfo{author}{\bibfnamefont{A.}~\bibnamefont{Carvalho}}, \bibnamefont{and}
  \bibinfo{author}{\bibfnamefont{A.~H.} \bibnamefont{Castro~Neto}},
  \bibinfo{journal}{Phys. Rev. Lett.} \textbf{\bibinfo{volume}{112}},
  \bibinfo{pages}{176801} (\bibinfo{year}{2014}).

\bibitem[{\citenamefont{{Rold{\'a}n} et~al.}(2015)\citenamefont{{Rold{\'a}n},
  {Castellanos-Gomez}, {Cappelluti}, and {Guinea}}}]{RG15}
\bibinfo{author}{\bibfnamefont{R.}~\bibnamefont{{Rold{\'a}n}}},
  \bibinfo{author}{\bibfnamefont{A.}~\bibnamefont{{Castellanos-Gomez}}},
  \bibinfo{author}{\bibfnamefont{E.}~\bibnamefont{{Cappelluti}}},
  \bibnamefont{and} \bibinfo{author}{\bibfnamefont{F.}~\bibnamefont{{Guinea}}},
  \bibinfo{journal}{J. Phys.: Condens. Matter} \textbf{\bibinfo{volume}{27}},
  \bibinfo{pages}{313201} (\bibinfo{year}{2015}).

\bibitem[{\citenamefont{Xiang et~al.}(2015)\citenamefont{Xiang, Ye, Shang, Lei,
  Wang, Yang, Liu, Meng, Luo, Zou et~al.}}]{XC15}
\bibinfo{author}{\bibfnamefont{Z.~J.} \bibnamefont{Xiang}},
  \bibinfo{author}{\bibfnamefont{G.~J.} \bibnamefont{Ye}},
  \bibinfo{author}{\bibfnamefont{C.}~\bibnamefont{Shang}},
  \bibinfo{author}{\bibfnamefont{B.}~\bibnamefont{Lei}},
  \bibinfo{author}{\bibfnamefont{N.~Z.} \bibnamefont{Wang}},
  \bibinfo{author}{\bibfnamefont{K.~S.} \bibnamefont{Yang}},
  \bibinfo{author}{\bibfnamefont{D.~Y.} \bibnamefont{Liu}},
  \bibinfo{author}{\bibfnamefont{F.~B.} \bibnamefont{Meng}},
  \bibinfo{author}{\bibfnamefont{X.~G.} \bibnamefont{Luo}},
  \bibinfo{author}{\bibfnamefont{L.~J.} \bibnamefont{Zou}},
  \bibnamefont{et~al.}, \bibinfo{journal}{Phys. Rev. Lett.}
  \textbf{\bibinfo{volume}{115}}, \bibinfo{pages}{186403}
  (\bibinfo{year}{2015}).

\bibitem[{\citenamefont{Quereda et~al.}(2016)\citenamefont{Quereda, San-Jose,
  Parente, Vaquero-Garzon, Molina-Mendoza, Agra•t, Rubio-Bollinger, Guinea,
  Rold‡n, and Castellanos-Gomez}}]{QC15}
\bibinfo{author}{\bibfnamefont{J.}~\bibnamefont{Quereda}},
  \bibinfo{author}{\bibfnamefont{P.}~\bibnamefont{San-Jose}},
  \bibinfo{author}{\bibfnamefont{V.}~\bibnamefont{Parente}},
  \bibinfo{author}{\bibfnamefont{L.}~\bibnamefont{Vaquero-Garzon}},
  \bibinfo{author}{\bibfnamefont{A.~J.} \bibnamefont{Molina-Mendoza}},
  \bibinfo{author}{\bibfnamefont{N.}~\bibnamefont{Agra•t}},
  \bibinfo{author}{\bibfnamefont{G.}~\bibnamefont{Rubio-Bollinger}},
  \bibinfo{author}{\bibfnamefont{F.}~\bibnamefont{Guinea}},
  \bibinfo{author}{\bibfnamefont{R.}~\bibnamefont{Rold‡n}}, \bibnamefont{and}
  \bibinfo{author}{\bibfnamefont{A.}~\bibnamefont{Castellanos-Gomez}},
  \bibinfo{journal}{Nano Letters} \textbf{\bibinfo{volume}{16}},
  \bibinfo{pages}{2931} (\bibinfo{year}{2016}).

\bibitem[{\citenamefont{Manjanath et~al.}(2015)\citenamefont{Manjanath,
  Samanta, Pandey, and Singh}}]{Manjanath2015}
\bibinfo{author}{\bibfnamefont{A.}~\bibnamefont{Manjanath}},
  \bibinfo{author}{\bibfnamefont{A.}~\bibnamefont{Samanta}},
  \bibinfo{author}{\bibfnamefont{T.}~\bibnamefont{Pandey}}, \bibnamefont{and}
  \bibinfo{author}{\bibfnamefont{A.~K.} \bibnamefont{Singh}},
  \bibinfo{journal}{Nanotechnology} \textbf{\bibinfo{volume}{26}},
  \bibinfo{pages}{075701} (\bibinfo{year}{2015}).

\bibitem[{\citenamefont{Jin et~al.}(2015)\citenamefont{Jin, Rold\'an,
  Katsnelson, and Yuan}}]{JRKY15}
\bibinfo{author}{\bibfnamefont{F.}~\bibnamefont{Jin}},
  \bibinfo{author}{\bibfnamefont{R.}~\bibnamefont{Rold\'an}},
  \bibinfo{author}{\bibfnamefont{M.~I.} \bibnamefont{Katsnelson}},
  \bibnamefont{and} \bibinfo{author}{\bibfnamefont{S.}~\bibnamefont{Yuan}},
  \bibinfo{journal}{Phys. Rev. B} \textbf{\bibinfo{volume}{92}},
  \bibinfo{pages}{115440} (\bibinfo{year}{2015}).

\bibitem[{\citenamefont{Rudenko and Katsnelson}(2014)}]{Rudenko2014}
\bibinfo{author}{\bibfnamefont{A.~N.} \bibnamefont{Rudenko}} \bibnamefont{and}
  \bibinfo{author}{\bibfnamefont{M.~I.} \bibnamefont{Katsnelson}},
  \bibinfo{journal}{Phys. Rev. B} \textbf{\bibinfo{volume}{89}},
  \bibinfo{pages}{201408} (\bibinfo{year}{2014}).

\bibitem[{\citenamefont{Rudenko et~al.}(2015)\citenamefont{Rudenko, Yuan, and
  Katsnelson}}]{Rudenko2015}
\bibinfo{author}{\bibfnamefont{A.~N.} \bibnamefont{Rudenko}},
  \bibinfo{author}{\bibfnamefont{S.}~\bibnamefont{Yuan}}, \bibnamefont{and}
  \bibinfo{author}{\bibfnamefont{M.~I.} \bibnamefont{Katsnelson}},
  \bibinfo{journal}{Phys. Rev. B} \textbf{\bibinfo{volume}{92}},
  \bibinfo{pages}{085419} (\bibinfo{year}{2015}).

\bibitem[{\citenamefont{Yuan et~al.}(2010)\citenamefont{Yuan, De~Raedt, and
  Katsnelson}}]{YRK10}
\bibinfo{author}{\bibfnamefont{S.}~\bibnamefont{Yuan}},
  \bibinfo{author}{\bibfnamefont{H.}~\bibnamefont{De~Raedt}}, \bibnamefont{and}
  \bibinfo{author}{\bibfnamefont{M.~I.} \bibnamefont{Katsnelson}},
  \bibinfo{journal}{Phys. Rev. B} \textbf{\bibinfo{volume}{82}},
  \bibinfo{pages}{115448} (\bibinfo{year}{2010}).

\bibitem[{\citenamefont{Yuan et~al.}(2011)\citenamefont{Yuan, Rold\'an, and
  Katsnelson}}]{Yuan2011}
\bibinfo{author}{\bibfnamefont{S.}~\bibnamefont{Yuan}},
  \bibinfo{author}{\bibfnamefont{R.}~\bibnamefont{Rold\'an}}, \bibnamefont{and}
  \bibinfo{author}{\bibfnamefont{M.~I.} \bibnamefont{Katsnelson}},
  \bibinfo{journal}{Phys. Rev. B} \textbf{\bibinfo{volume}{84}},
  \bibinfo{pages}{035439} (\bibinfo{year}{2011}).

\bibitem[{\citenamefont{Yuan et~al.}(2012{\natexlab{a}})\citenamefont{Yuan,
  Wehling, Lichtenstein, and Katsnelson}}]{Yuan2012}
\bibinfo{author}{\bibfnamefont{S.}~\bibnamefont{Yuan}},
  \bibinfo{author}{\bibfnamefont{T.~O.} \bibnamefont{Wehling}},
  \bibinfo{author}{\bibfnamefont{A.~I.} \bibnamefont{Lichtenstein}},
  \bibnamefont{and} \bibinfo{author}{\bibfnamefont{M.~I.}
  \bibnamefont{Katsnelson}}, \bibinfo{journal}{Phys. Rev. Lett.}
  \textbf{\bibinfo{volume}{109}}, \bibinfo{pages}{156601}
  (\bibinfo{year}{2012}{\natexlab{a}}).

\bibitem[{\citenamefont{Pereira and Katsnelson}(2015)}]{Pereira15}
\bibinfo{author}{\bibfnamefont{J.~M.} \bibnamefont{Pereira}} \bibnamefont{and}
  \bibinfo{author}{\bibfnamefont{M.~I.} \bibnamefont{Katsnelson}},
  \bibinfo{journal}{Phys. Rev. B} \textbf{\bibinfo{volume}{92}},
  \bibinfo{pages}{075437} (\bibinfo{year}{2015}).

\bibitem[{\citenamefont{Jiang et~al.}(2015)\citenamefont{Jiang, Rold\'an,
  Guinea, and Low}}]{Jiang2015}
\bibinfo{author}{\bibfnamefont{Y.}~\bibnamefont{Jiang}},
  \bibinfo{author}{\bibfnamefont{R.}~\bibnamefont{Rold\'an}},
  \bibinfo{author}{\bibfnamefont{F.}~\bibnamefont{Guinea}}, \bibnamefont{and}
  \bibinfo{author}{\bibfnamefont{T.}~\bibnamefont{Low}},
  \bibinfo{journal}{Phys. Rev. B} \textbf{\bibinfo{volume}{92}},
  \bibinfo{pages}{085408} (\bibinfo{year}{2015}).

\bibitem[{\citenamefont{Zhou et~al.}(2015)\citenamefont{Zhou, Zhang, Sun, Zou,
  Zhou, Zhai, and Chang}}]{zhou15}
\bibinfo{author}{\bibfnamefont{X.}~\bibnamefont{Zhou}},
  \bibinfo{author}{\bibfnamefont{R.}~\bibnamefont{Zhang}},
  \bibinfo{author}{\bibfnamefont{J.}~\bibnamefont{Sun}},
  \bibinfo{author}{\bibfnamefont{Y.}~\bibnamefont{Zou}},
  \bibinfo{author}{\bibfnamefont{G.}~\bibnamefont{Zhou}},
  \bibinfo{author}{\bibfnamefont{F.}~\bibnamefont{Zhai}}, \bibnamefont{and}
  \bibinfo{author}{\bibfnamefont{K.}~\bibnamefont{Chang}},
  \bibinfo{journal}{Scientific Reports} \textbf{\bibinfo{volume}{5}},
  \bibinfo{pages}{12295} (\bibinfo{year}{2015}).

\bibitem[{\citenamefont{Bastin et~al.}(1971)\citenamefont{Bastin, Lewiner,
  Betbeder-matibet, and Nozieres}}]{Bastin71}
\bibinfo{author}{\bibfnamefont{A.}~\bibnamefont{Bastin}},
  \bibinfo{author}{\bibfnamefont{C.}~\bibnamefont{Lewiner}},
  \bibinfo{author}{\bibfnamefont{O.}~\bibnamefont{Betbeder-matibet}},
  \bibnamefont{and} \bibinfo{author}{\bibfnamefont{P.}~\bibnamefont{Nozieres}},
  \bibinfo{journal}{Journal of Physics and Chemistry of Solids}
  \textbf{\bibinfo{volume}{32}}, \bibinfo{pages}{1811 } (\bibinfo{year}{1971}).

\bibitem[{\citenamefont{Garc\'{\i}a et~al.}(2015)\citenamefont{Garc\'{\i}a,
  Covaci, and Rappoport}}]{GCR15}
\bibinfo{author}{\bibfnamefont{J.~H.} \bibnamefont{Garc\'{\i}a}},
  \bibinfo{author}{\bibfnamefont{L.}~\bibnamefont{Covaci}}, \bibnamefont{and}
  \bibinfo{author}{\bibfnamefont{T.~G.} \bibnamefont{Rappoport}},
  \bibinfo{journal}{Phys. Rev. Lett.} \textbf{\bibinfo{volume}{114}},
  \bibinfo{pages}{116602} (\bibinfo{year}{2015}).

\bibitem[{\citenamefont{Ortmann et~al.}(2015)\citenamefont{Ortmann, Leconte,
  and Roche}}]{Ortmann2015}
\bibinfo{author}{\bibfnamefont{F.}~\bibnamefont{Ortmann}},
  \bibinfo{author}{\bibfnamefont{N.}~\bibnamefont{Leconte}}, \bibnamefont{and}
  \bibinfo{author}{\bibfnamefont{S.}~\bibnamefont{Roche}},
  \bibinfo{journal}{Phys. Rev. B} \textbf{\bibinfo{volume}{91}},
  \bibinfo{pages}{165117} (\bibinfo{year}{2015}).

\bibitem[{\citenamefont{Goerbig}(2011)}]{G11}
\bibinfo{author}{\bibfnamefont{M.}~\bibnamefont{Goerbig}},
  \bibinfo{journal}{Reviews of Modern Physics} \textbf{\bibinfo{volume}{83}},
  \bibinfo{pages}{1193} (\bibinfo{year}{2011}).

\bibitem[{\citenamefont{Hams and De~Raedt}(2000)}]{HR00}
\bibinfo{author}{\bibfnamefont{A.}~\bibnamefont{Hams}} \bibnamefont{and}
  \bibinfo{author}{\bibfnamefont{H.}~\bibnamefont{De~Raedt}},
  \bibinfo{journal}{Phys. Rev. E} \textbf{\bibinfo{volume}{62}},
  \bibinfo{pages}{4365} (\bibinfo{year}{2000}).

\bibitem[{\citenamefont{Dietl et~al.}(2008)\citenamefont{Dietl, Pi\'echon, and
  Montambaux}}]{DPM08}
\bibinfo{author}{\bibfnamefont{P.}~\bibnamefont{Dietl}},
  \bibinfo{author}{\bibfnamefont{F.}~\bibnamefont{Pi\'echon}},
  \bibnamefont{and}
  \bibinfo{author}{\bibfnamefont{G.}~\bibnamefont{Montambaux}},
  \bibinfo{journal}{Phys. Rev. Lett.} \textbf{\bibinfo{volume}{100}},
  \bibinfo{pages}{236405} (\bibinfo{year}{2008}).

\bibitem[{\citenamefont{Montambaux
  et~al.}(2009{\natexlab{a}})\citenamefont{Montambaux, Piéchon, Fuchs, and
  Goerbig}}]{MG09b}
\bibinfo{author}{\bibfnamefont{G.}~\bibnamefont{Montambaux}},
  \bibinfo{author}{\bibfnamefont{F.}~\bibnamefont{Piéchon}},
  \bibinfo{author}{\bibfnamefont{J.-N.} \bibnamefont{Fuchs}}, \bibnamefont{and}
  \bibinfo{author}{\bibfnamefont{M.~O.} \bibnamefont{Goerbig}},
  \bibinfo{journal}{The European Physical Journal B}
  \textbf{\bibinfo{volume}{72}}, \bibinfo{pages}{509}
  (\bibinfo{year}{2009}{\natexlab{a}}).

\bibitem[{\citenamefont{Montambaux
  et~al.}(2009{\natexlab{b}})\citenamefont{Montambaux, Pi\'echon, Fuchs, and
  Goerbig}}]{MG09}
\bibinfo{author}{\bibfnamefont{G.}~\bibnamefont{Montambaux}},
  \bibinfo{author}{\bibfnamefont{F.}~\bibnamefont{Pi\'echon}},
  \bibinfo{author}{\bibfnamefont{J.-N.} \bibnamefont{Fuchs}}, \bibnamefont{and}
  \bibinfo{author}{\bibfnamefont{M.~O.} \bibnamefont{Goerbig}},
  \bibinfo{journal}{Phys. Rev. B} \textbf{\bibinfo{volume}{80}},
  \bibinfo{pages}{153412} (\bibinfo{year}{2009}{\natexlab{b}}).

\bibitem[{\citenamefont{Fuchs et~al.}(2010)\citenamefont{Fuchs, Pi{\'e}chon,
  Goerbig, and Montambaux}}]{FM10}
\bibinfo{author}{\bibfnamefont{J.}~\bibnamefont{Fuchs}},
  \bibinfo{author}{\bibfnamefont{F.}~\bibnamefont{Pi{\'e}chon}},
  \bibinfo{author}{\bibfnamefont{M.}~\bibnamefont{Goerbig}}, \bibnamefont{and}
  \bibinfo{author}{\bibfnamefont{G.}~\bibnamefont{Montambaux}},
  \bibinfo{journal}{The European Physical Journal B}
  \textbf{\bibinfo{volume}{77}}, \bibinfo{pages}{351} (\bibinfo{year}{2010}).

\bibitem[{\citenamefont{de~Gail et~al.}(2012)\citenamefont{de~Gail, Goerbig,
  and Montambaux}}]{GGM12}
\bibinfo{author}{\bibfnamefont{R.}~\bibnamefont{de~Gail}},
  \bibinfo{author}{\bibfnamefont{M.~O.} \bibnamefont{Goerbig}},
  \bibnamefont{and}
  \bibinfo{author}{\bibfnamefont{G.}~\bibnamefont{Montambaux}},
  \bibinfo{journal}{Phys. Rev. B} \textbf{\bibinfo{volume}{86}},
  \bibinfo{pages}{045407} (\bibinfo{year}{2012}).

\bibitem[{\citenamefont{Hatsugai et~al.}(2006)\citenamefont{Hatsugai, Fukui,
  and Aoki}}]{Hatsugai2006}
\bibinfo{author}{\bibfnamefont{Y.}~\bibnamefont{Hatsugai}},
  \bibinfo{author}{\bibfnamefont{T.}~\bibnamefont{Fukui}}, \bibnamefont{and}
  \bibinfo{author}{\bibfnamefont{H.}~\bibnamefont{Aoki}},
  \bibinfo{journal}{Phys. Rev. B} \textbf{\bibinfo{volume}{74}},
  \bibinfo{pages}{205414} (\bibinfo{year}{2006}).

\bibitem[{\citenamefont{Yuan et~al.}(2012{\natexlab{b}})\citenamefont{Yuan,
  Rold{\'a}n, and Katsnelson}}]{YRK12}
\bibinfo{author}{\bibfnamefont{S.}~\bibnamefont{Yuan}},
  \bibinfo{author}{\bibfnamefont{R.}~\bibnamefont{Rold{\'a}n}},
  \bibnamefont{and} \bibinfo{author}{\bibfnamefont{M.~I.}
  \bibnamefont{Katsnelson}}, \bibinfo{journal}{Solid State Communications}
  \textbf{\bibinfo{volume}{152}}, \bibinfo{pages}{1446}
  (\bibinfo{year}{2012}{\natexlab{b}}).

\bibitem[{\citenamefont{Wei\ss{}e et~al.}(2006)\citenamefont{Wei\ss{}e,
  Wellein, Alvermann, and Fehske}}]{KPM}
\bibinfo{author}{\bibfnamefont{A.}~\bibnamefont{Wei\ss{}e}},
  \bibinfo{author}{\bibfnamefont{G.}~\bibnamefont{Wellein}},
  \bibinfo{author}{\bibfnamefont{A.}~\bibnamefont{Alvermann}},
  \bibnamefont{and} \bibinfo{author}{\bibfnamefont{H.}~\bibnamefont{Fehske}},
  \bibinfo{journal}{Rev. Mod. Phys.} \textbf{\bibinfo{volume}{78}},
  \bibinfo{pages}{275} (\bibinfo{year}{2006}).

\bibitem[{\citenamefont{Mayorov et~al.}(2011)\citenamefont{Mayorov, Elias,
  Mucha-Kruczynski, Gorbachev, Tudorovskiy, Zhukov, Morozov, Katsnelson,
  Fal?ko, Geim et~al.}}]{Mayorov_Science2011}
\bibinfo{author}{\bibfnamefont{A.}~\bibnamefont{Mayorov}},
  \bibinfo{author}{\bibfnamefont{D.}~\bibnamefont{Elias}},
  \bibinfo{author}{\bibfnamefont{M.}~\bibnamefont{Mucha-Kruczynski}},
  \bibinfo{author}{\bibfnamefont{R.}~\bibnamefont{Gorbachev}},
  \bibinfo{author}{\bibfnamefont{T.}~\bibnamefont{Tudorovskiy}},
  \bibinfo{author}{\bibfnamefont{A.}~\bibnamefont{Zhukov}},
  \bibinfo{author}{\bibfnamefont{S.}~\bibnamefont{Morozov}},
  \bibinfo{author}{\bibfnamefont{M.}~\bibnamefont{Katsnelson}},
  \bibinfo{author}{\bibfnamefont{V.}~\bibnamefont{Fal?ko}},
  \bibinfo{author}{\bibfnamefont{A.}~\bibnamefont{Geim}}, \bibnamefont{et~al.},
  \bibinfo{journal}{Science} \textbf{\bibinfo{volume}{333}},
  \bibinfo{pages}{860} (\bibinfo{year}{2011}).

\end{thebibliography}
\end{document}